\UseRawInputEncoding 

\documentclass[aps,showpacs,onecolumn,notitlepage,floatfix,amsmath,amssymb,nofootinbib]{revtex4-1}

\usepackage{amssymb}
\usepackage{amsmath}
\usepackage{epsfig}
\usepackage{graphicx}
\usepackage{subfig}
\usepackage{color}
\usepackage{bigints}
\usepackage{latexsym}
\usepackage{bm,latexsym}
\usepackage{mathrsfs}
\usepackage{ulem}
\usepackage{float}
\usepackage{pbox}
\usepackage[usenames,dvipsnames]{xcolor}
\definecolor{orange}{cmyk}{0,0.5,1,0}

\selectfont{}

\begin{document}

\title{ Simple method to generate magnetically charged ultrastatic traversable wormholes without exotic 
matter in Einstein-scalar-Gauss-Bonnet  gravity }

\author{Pedro Ca\~nate} 
\email[]{pcannate@gmail.com }
\affiliation{Programa de F\'isica, Facultad de Ciencias Exactas y Naturales, Universidad Surcolombiana, Avenida Pastrana Borrero - Carrera 1, A.A. 385, C.P. 410001, Neiva, Huila, Colombia }

\begin{abstract}
All the magnetically charged ultrastatic and spherically symmetric  spacetime solutions in the framework of linear/nonlinear electrodynamics, with an arbitrary electromagnetic Lagrangian density $\mathcal{L}(\mathcal{F})$ depending only of the electromagnetic invariant $\mathcal{F}\!=\!F_{\alpha\beta}F^{\alpha\beta}\!/4$, 
minimally  coupled to Einstein-scalar-Gauss-Bonnet gravity  [EsGB-$\mathcal{L}(\mathcal{F})$], are found. 
We also show that a magnetically charged ultrastatic and spherically symmetric 
EsGB-$\mathcal{L}(\mathcal{F})$ solution with invariant $\mathcal{F}$ having a strict global  maximum value $\mathcal{F}_{_{0}}$ in the entire domain of the solution, and  such that $\mathcal{L}_{_{0}}=\mathcal{L}(\mathcal{F}_{_{0}})>0$, can be interpreted as an ultrastatic wormhole  spacetime geometry  with throat radius determined by the scalar charge and the quantity $\mathcal{L}_{_{0}}$. 
We provide some examples, including Maxwell's theory of electrodynamics (linear electrodynamics)   
$\mathcal{L}_{_{_{\mathrm{LED}}}} \!=\! \mathcal{F}$, producing the magnetic dual of the purely electric Ellis-Bronnikov EsGB Maxwell wormhole derived in 
[P. Ca\~nate, J. Sultana, D. Kazanas, Phys. Rev. D {\bf100}, 064007 (2019)];   
and the nonlinear electrodynamics (NLED) models given by
Born-Infeld $\mathcal{L}_{_{_{\mathrm{BI}}}} \!=\! -4\beta^{2} + 4\beta^{2} \sqrt{ 1 + \mathcal{F}\!/\!(2\beta^{2})~}$, and Euler-Heisenberg in the approximation of the weak-field limit $\mathcal{L}_{_{_{\mathrm{EH}}}} \!=\! \mathcal{L}_{_{_{\mathrm{LED}}}} + \gamma \mathcal{F}^{2}\!/2$.  
With those NLED models, two novel magnetically charged ultrastatic traversable wormholes (EsGB Born-Infeld and EsGB Euler-Heisenberg wormholes) are presented as exact solutions without exotic matter 
in EsGB-$\mathcal{L}(\mathcal{F})$ gravity, and we show that these solutions  
have in common the property that in the weak electromagnetic field region  
the magnetically charged Ellis-Bronnikov EsGB Maxwell wormhole is recuperated. 
\end{abstract}



\maketitle

\section{Introduction} 
Wormholes are an interesting type of spacetimes which arises from the geometrical description of gravity
\cite{visser95}. These involve a topological spacetime configuration in the form of a shortcut that links two spacetimes or two distinct regions of the same spacetime.
The idea originates from the work of Einstein and Rosen in 1935, with their solution known as the Einstein-Rosen bridge \cite{einstein35}, which is basically the maximally extended Schwarzschild solution. However, it quickly turned out that the ``throat" of such a wormhole is dynamic and hence nontraversable \cite{kruskal60}, meaning that its radius expands to a maximum and quickly contracts to zero so fast that even a photon cannot pass through. Following this, interest in wormholes was revived by the seminal work of Morris and Thorne in 1988 \cite{morris88}, who discuss the construction of
traversable wormholes (T-WHs) within the general relativity (GR) context, and showed that the throat of these wormholes can be kept open by some form of  ``exotic" matter \cite{morris88-2} having negative energy density, and whose energy momentum tensor violates the null-energy condition (NEC).  
This suggests that, at a classical level, T-WHs of the Morris-Thorne type 
are forbidden in general relativity because all know types of physically reasonable matter satisfy the NEC. 

Nevertheless, at the quantum level the need for NEC violation is not in itself a big problem; quantum fields can easily violate NEC \cite{Epstein65}.
From this perspective, recently,  using ideas from gauge/gravity duality, it has been shown that quantum matter fields can provide the necessary negative energy to keep the throat of the wormhole open and thus achieve traversable wormholes (see \cite{gao17,fu19} for
a review). 
More recently \cite{horowitz19}, it was shown that wormholes can also be produced through a quantum tunneling event, where a backreaction from quantum fields can make these wormholes traversable.

Lately, in order to avoid the exotic matter issue in the traversable Morris-Thorne wormholes at a classical level,
several modifications to GR have been explored,\footnote{ Modified theories of gravity have been developed primarily in the hope of finding solutions to the several issues of current observational astrophysics and cosmology (such as dark matter and dark energy questions).} for instance see 
\cite{duplessis15,nascimento17,bronnikov16,lobo08,zangeneh15,ovgun19,zubair18,bohemer12,evseev18,mironov18,Harko14}.
More precisely, the necessity of the presence of exotic matter in 
classical
T-WHs can be circumvented in modified theories of gravity which contain higher curvature corrections in their gravitational actions, so that
additional curvature degrees of freedom can support such geometries. 
In order to do so, it is necessary that 
the effect of the higher-order curvature terms leads to the existence of spacetime regions with negative effective energy density and can thus mimic the exotic matter contribution required for T-WHs. 

It should be pointed out that the most natural higher-order curvature extension of $D$-dimensional general relativity,  
satisfying the criteria of general covariance and leading to second order field equations for the metric, is given by the Lanczos-Lovelock (LL) gravity theory \cite{Lovelock},  which is defined by the action
\begin{equation}\label{actionLovelock}
S_{_{\rm LL}}[g_{\mu\nu}, \boldsymbol{\psi}] = \int d^{D}x \sqrt{-g}\left( \frac{1}{16\pi} \sum_{p=0}^{\ell} \alpha_{_{(p)}}\mathcal{L}_{_{(p)}}(R, R_{\mu\nu},R_{\mu\nu\sigma\beta}) \right) + \mathcal{S}_{_{\rm matter}}[g_{\mu\nu}, \boldsymbol{\psi}] 
\end{equation}
where the first term to the right side of the equation
is the Lovelock action for $D$-dimensional spacetimes with metric signature $(-,+,+\!~,\!~.\!~.\!~.\!~,\!~+)$, while the second is the usual action for the matter, where $\boldsymbol{\psi}$ represents schematically the matter fields.
Here, $\mathcal{L}_{_{(p)}}$ are functions of the curvature scalar $R$, and the curvature tensors $R_{\mu\nu}$, $R_{\mu\nu\alpha\beta}$, given by
\begin{equation}\label{DensLovelock}
\mathcal{L}_{_{(0)}} = 1 \quad\quad\textup{ and }\quad\quad \mathcal{L}_{_{(p)}} = \frac{1}{2^{^{p}}}\delta^{^{\mu_{_{1}}\nu_{_{1}}\mu_{_{2}}\nu_{_{2}}\cdot\cdot\cdot\mu_{_{p}}\nu_{_{p}}}}_{_{\beta_{_{1}}\sigma_{_{1}}\beta_{_{2}}\sigma_{_{2}}\cdot\cdot\cdot\beta_{_{p}}\sigma_{_{p}}}}\prod_{j=1}^{p}R_{\mu_{_{j}}\nu_{_{j}}}{}^{\beta_{_{j}}\sigma_{_{j}}} \quad\quad\forall p\neq 0
\end{equation}
where $\delta^{^{\mu_{_{1}}\nu_{_{1}}\mu_{_{2}}\nu_{_{2}}\cdot\cdot\cdot\mu_{_{p}}\nu_{_{p}}}}_{_{\beta_{_{1}}\sigma_{_{1}}\beta_{_{2}}\sigma_{_{2}}\cdot\cdot\cdot\beta_{_{p}}\sigma_{_{p}}}}$ is the generalized totally antisymmetric Kronecker delta,\footnote{ The generalized totally antisymmetric Kronecker delta is defined as 
\begin{equation}
\delta^{^{\mu_{_{1}}\nu_{_{1}}\mu_{_{2}}\nu_{_{2}}\cdot\cdot\cdot\mu_{_{p}}\nu_{_{p}}}}_{_{\beta_{_{1}}\sigma_{_{1}}\beta_{_{2}}\sigma_{_{2}}\cdot\cdot\cdot\beta_{_{p}}\sigma_{_{p}}}} = (2p)! \delta^{^{[\mu_{_{1}}}}_{_{\beta_{_{1}}}} \delta^{^{\nu_{_{1}}}}_{_{\sigma_{_{1}}}} \delta^{^{\mu_{_{2}}}}_{_{\beta_{_{2}}}} \delta^{^{\nu_{_{2}}}}_{_{\sigma_{_{2}}}} \cdot\cdot\cdot \delta^{^{\mu_{_{p}}}}_{_{\beta_{_{p}}}} \delta^{^{\nu_{_{p}}]}}_{_{\sigma_{_{p}}}} = (2p)! \delta^{^{\mu_{_{1}}}}_{_{[\beta_{_{1}}}} \delta^{^{\nu_{_{1}}}}_{_{\sigma_{_{1}}}} \delta^{^{\mu_{_{2}}}}_{_{\beta_{_{2}}}} \delta^{^{\nu_{_{2}}}}_{_{\sigma_{_{2}}}} \cdot\cdot\cdot \delta^{^{\mu_{_{p}}}}_{_{\beta_{_{p}}}} \delta^{^{\nu_{_{p}}}}_{_{\sigma_{_{p}}]}}    
\end{equation}
} the parameters $\alpha_{_{(p)}}$ are coupling constants, being $\alpha_{_{(0)}} = -2\Lambda$ (cosmological constant), $\alpha_{_{(1)}}=1$, whereas $\{\alpha_{_{(j)}}\}_{_{j=2}}^{^{\ell}}$
are arbitrary constants of the theory,  
and the parameter $\ell$ is called the order of the Lovelock gravity.
Remarkably, the Lovelock gravity besides being special because its field equations are still second-order differential equations for the metric tensor components $g_{\mu\nu}$ (as in GR), 
this theory has also gained much importance because it appears as
the low energy limit of string theories, where the parameter $\alpha_{_{(2)}}$ (known as the second order Lovelock coefficient) is proportional to the inverse string tension \cite{Boulware}, or related to the coupling constant $\alpha'$ in the string world-sheet action \cite{Barton85}. 

Let us now sketch some reductions of the action \eqref{actionLovelock}. 
First, notice that the generalized totally antisymmetric Kronecker delta in Eq. \eqref{DensLovelock} causes  
$\mathcal{L}_{_{(p)}}$ to become identically zero for all $p > [D/2]$ where $[D/2]$ denotes the integral part of $D/2$, and then w.l.o.g one can always set $\ell=[D/2]$. 
The $\mathcal{L}_{_{(p)}}$ functions for $p=1,2,3$ are    
$\mathcal{L}_{_{(1)}} = R$, $\mathcal{L}_{_{(2)}} = R^{2} - 4R_{\mu\nu}R^{\mu\nu} + R_{\mu\nu\sigma\rho}R^{\mu\nu\sigma\rho}$ and  
\begin{eqnarray}
&&\mathcal{L}_{_{(3)}} = R^{3} - 12RR_{\mu\nu}R^{\mu\nu} + 16R_{\mu\nu}R^{\mu}{}_{\sigma}R^{\nu\sigma} + 24R_{\mu\nu}R_{\sigma\rho}R^{\mu\sigma\nu\rho} + 3RR_{\mu\nu\sigma\rho}R^{\mu\nu\sigma\rho} - 24R_{\mu\nu}R^{\mu}{}_{\sigma\rho\gamma}R^{\nu\sigma\rho\gamma} \nonumber\\
&& \quad\quad + \quad\!\!\!\!4 R_{\mu\nu\sigma\rho}R^{\mu\nu\gamma\eta}R^{\sigma\rho}{}_{\gamma\eta}-8R_{\mu\nu\sigma\rho}R^{\mu}{}_{\gamma}{}^{\sigma}{}_{\eta}R^{\nu\gamma\rho\eta}. 
\end{eqnarray} 
{\it First ($\ell$=1)- and second ($\ell$=2)-order curvature LL gravity.} Expanding the Lovelock action up to $\ell=1$ (for an arbitrary value of $D$), the $D$-dimensional Einstein-Hilbert action  with cosmological constant  is recovered; whereas, working up to $\ell=2$ yields the $D$-dimensional Einstein-Gauss-Bonnet (EGB) theory with cosmological constant, which is defined by the action  
\begin{equation}\label{actionGB}
S_{_{\rm EGB}}[g_{\mu\nu}, \boldsymbol{\psi}] = \int d^{^{D}}\!\!x \sqrt{-g} \left\{ \frac{1}{16\pi}\left(R - 2\Lambda + \alpha_{_{GB}} R_{_{GB}}^{2}\right) \right\} + \mathcal{S}_{_{\rm matter}}[g_{\mu\nu}, \boldsymbol{\psi}]
\end{equation}
where $R_{_{GB}}^{2}$ stands for the quadratic Gauss-Bonnet (GB) term (also known as Gauss-Bonnet invariant) defined by $R_{_{GB}}^{2} = R_{\sigma\beta\mu\nu}R^{\sigma\beta\mu\nu} - 4R_{\sigma\beta}R^{\sigma\beta} + R^{2}$; 
while $\alpha_{_{GB}}$ is a dimensionless coupling constant
known as the Gauss-Bonnet coefficient.   
So that, according with Eq. \eqref{actionLovelock}, $\mathcal{L}_{_{(0)}} + \mathcal{L}_{_{(1)}}$ corresponds to the Einstein-Hilbert Lagrangian density with cosmological constant,  
while $\mathcal{L}_{_{(2)}}$ is the quadratic Gauss-Bonnet term $\mathcal{L}_{_{(2)}}=R_{_{GB}}^{2}$ and $\alpha_{_{(2)}}=\alpha_{_{GB}}$.
In this context, for $D\geq 5$, $D$-dimensional Lorentzian wormhole\footnote{That is, higher-dimensional T-WHs of the Morris-Thorne type.} solutions are investigated in the Ref. 
\cite{WH_5EGB}. 

It is worth stressing that in $D = 4$ the  quadratic curvature term $\alpha_{_{GB}}R_{_{GB}}^{2}$ (with $\alpha_{_{GB}}$ finite) is a total derivative, and hence does not contribute to the gravitational dynamics. 
In fact, the Gauss-Bonnet term contribution to all the components of modified Einstein's field equations  
are proportional to $(D-4)$,
and therefore vanish identically at $D=4$,
see \cite{Mardones1991,Torii2008} for a discussion; 
whereas, in $D=4$ with singular $\alpha_{_{GB}}$  (i.e. $\alpha_{_{GB}} = \frac{\tilde{\alpha}_{_{GB}}}{D-4}$, being $\tilde{\alpha}_{_{GB}}\in\mathbb{R}$) yields a nontrivial case recently 
discussed in \cite{Glavan2020,Gurses2020}.

Summarizing, we can say that in four-dimensional gravity the quadratic curvature term, $\alpha_{_{GB}}R_{_{GB}}^{2}$ (with regular $\alpha_{_{GB}}$), does not modify the Einstein field equations.\footnote{The action \eqref{actionGB} leads to the following modified Einstein's field equations in $D$-dimensional spacetime: 
\begin{equation}
G_{\mu\nu} + \alpha_{_{_{GB}}} \mathcal{H}_{\mu\nu} + g_{\mu\nu} \Lambda =  8\pi T_{\mu\nu}
\end{equation}
where $G_{\nu\beta}$ and $T_{\nu\beta}$, respectively, are  
the components of Einstein tensor and energy-momentum tensor of the matter fields; whereas 
\begin{equation}
\mathcal{H}_{\mu\nu} =  2\left( RR_{\mu\nu} - 2R_{\mu\sigma\nu\beta}R^{\sigma\beta} + R_{\mu\sigma\beta\gamma}R_{\nu}{}^{\sigma\beta\gamma} - 2R_{\mu\sigma}R_{\nu}{}^{\sigma}  \right) - \frac{1}{2} g_{\mu\nu}R_{_{GB}}^{2}. 
\end{equation}
In particular, $\mathcal{H}_{\mu\nu}$ vanishes identically in $D = 4$. } 
However, if a massless scalar field, $\phi$, is coupled to the GB term through a nontrivial well-defined coupling function\footnote{I.e., $\boldsymbol{f}(\phi)\neq$ constant, and $\boldsymbol{f}(\phi)\in\mathcal{C}^{2}$.} $\boldsymbol{f}(\phi)$, then the new field equations are substantially different from the standard general relativity due to the presence of the $\boldsymbol{f}(\phi)$-Gauss-Bonnet ($\boldsymbol{f}$GB) curvature tensor (which will be introduced later). 
In this way arises one of the best four-dimensional higher curvature gravity theories, known as the Einstein-scalar-Gauss-Bonnet (EsGB) theory.\footnote{Also referred to as generalized  Einstein-dilaton-Gauss-Bonnet  gravity.} 
The general action for an EsGB theory in the presence of a cosmological constant $\Lambda$, with an arbitrary coupling function $\boldsymbol{f}(\phi)$ between the scalar field and the GB term, is given by
\begin{equation}\label{actionLEsGB}
S_{_{\rm EsGB}}[g_{\mu\nu}, \phi, \boldsymbol{\psi}] = \int d^{4}x \sqrt{-g} \left\{ \frac{1}{16\pi}\left(R - 2\Lambda - \frac{1}{2}\partial_{\alpha}\phi\partial^{\alpha}\phi + \boldsymbol{f}(\phi) R_{_{GB}}^{2}  \right)   
\right\} + \mathcal{S}_{_{\rm matter}}[g_{\mu\nu}, \boldsymbol{\psi}].
\end{equation} 
Despite the presence of a quadratic curvature invariant $\boldsymbol{f}(\phi)R_{_{GB}}^{2}$ in the  EsGB action, the resulting field equations arising from the variation of this action with respect to the metric are of second order and therefore avoid the Ostrogradski instability and ghosts.

In similarity with LL theory,  the EsGB gravity also arises from the low energy limit of a
heterotic string theory  \cite{Metsaev87}.  Furthermore, the EsGB gravity can be interpreted as a subclass of Horndeski gravity \cite{Kobayashi11,Kobayashi19} and also arise from an effective field theory perspective \cite{Yagi16,Cano20}.  
As a consequence of the above aspects, the predictions of the EsGB theory are dramatically different
from GR in high-curvature regions, such as the early universe epochs, and the interior of black holes, where it aims at resolving curvature singularities. Indeed, the 
coupling between the GB term and the scalar field is responsible for the fact that the black holes in the EsGB gravity context can violate no-hair theorems \cite{Bekenstein72,Sudarsky95} and exhibit spontaneous and induced scalarization of black holes arise; the stability of those scalarized black holes has been addressed in \cite{Kunz2018}.
Also, free curvature singularity black holes (that evade the Penrose singularity theorem \cite{Penrose}) have been numerically derived in EsGB, see \cite{Kanti2018}. 
It is also noticeable that the Gauss-Bonnet curvature tensor  
allows the presence of regions with negative effective energy densities \cite{Kunz2011}, and one of the consequences is that structures like T-WHs can be sustained without invoking exotic matter.
For instance, in \cite{Kanti2011} linearly stable T-WHs 
were numerically derived with a coupling function given by $f(\phi)= e^{- \gamma \phi}$ with no need  of exotic matter for supporting the wormhole. 

Recently \cite{antoniou19}, a number of novel T-WH 
solutions in EsGB theory have been obtained numerically for several coupling functions. Other numerically 
T-WHs have been obtained earlier in Einstein-dilaton-Gauss-Bonnet theory \cite{kanti11}, which involve an exponential coupling between the scalar field representing the dilaton and the Gauss-Bonnet term.\\ 
The construction of interesting exact solutions like black holes or novel T-WHs  in the framework of the EsGB gravity, to date, has only been possible using nonlinear electromagnetic fields as sources in the EsGB field equations (see Refs. \cite{Canate2020,Canate2019,canate19b,Canate2022}).
Concretely, the nonlinear electrodynamics (NLED) theories are extensions of Maxwell's electromagnetism
that suggest that the Lagrangian density of the electromagnetic field  depends in a nonlinear way on the two electromagnetic invariants, $\mathcal{F}= 2(\boldsymbol{\mathcal{B}}^2-\boldsymbol{\mathcal{E}}^2)$ and $\mathcal{G} = \boldsymbol{\mathcal{E}}\cdot\boldsymbol{\mathcal{B}}$,  where $\boldsymbol{\mathcal{E}}$ and $\boldsymbol{\mathcal{B}}$ are the electric and magnetic fields, respectively. Therefore, the most general NLED Lagrangian density is  
characterized by an arbitrary function $\mathcal{L}(\mathcal{F}, \mathcal{G})$ of the electromagnetic invariants, $\mathcal{F}$ and $\mathcal{G}$.  
For more details on these aspects see \cite{BI,JFP70,Salazar87,Heuler_35,Heisenberg_36} (see also \cite{Sorokin21} for a recent review) and references within.
The most simple $\mathcal{L}(\mathcal{F}, \mathcal{G})$ model 
corresponds to the linear case (LED), given by
\begin{equation}\label{Maxw}
\mathcal{L}_{_{_{\mathrm{LED}}}} 
= \mathcal{F} 
\end{equation}
which is known as Maxwell's theory of electrodynamics.\footnote{
The  Maxwell's electrodynamics is one of the most notable and experimentally verified classical field theories 
ever constructed.  Since its formulation (about 1860), it has been the source of remarkable predictions such as the electromagnetic radiation.
In addition, the Maxwell's theory has served as a keystone for the proposal of new theories, such as Einstein's theory of special relativity.}
In 1933$-$1934 Born and Infeld (BI) constructed the first nonlinear generalization of Maxwell's electrodynamics for strong fields, which was invented to ensure that electric field self-energy of charged point particles is finite and, therefore, a solution of the Maxwell's electrodynamics problem of point charges and their diverging self-energy is proposed \cite{BI}.
The proposed BI Lagrangian that depends nonlinearly on the electromagnetic invariants $\mathcal{F}$ and $\mathcal{G}$ was inspired in a finiteness principle for the electromagnetic field magnitude (analogous to the special relativity theory that assumed an upper limit to the velocity of light), and has the form
\begin{equation}\label{BI2}
\mathcal{L}_{_{\mathrm{BI}}}  
= 4\beta^{2} \left( -1 + \sqrt{ 1 + \frac{\mathcal{F}}{2\beta^{2}} + \frac{\mathcal{G}^{2}}{16\beta^{4}} 
}~\right),
\end{equation}
where $\beta$ is a constant which has the physical interpretation of a critical field strength. 
Later, in 1935  Euler and Heisenberg (EH) computed a complete effective action describing nonlinear corrections to Maxwell's theory due to quantum electron-positron one-loop effects. In a nonperturbative
form, it is given by
\begin{equation}\label{EH}
\mathcal{L}_{_{\mathrm{EH}}}  
= \mathcal{L}_{_{_{\mathrm{LED}}}} + \frac{1}{8\pi^{2}}\int^{^{\infty}}_{_{0}} \frac{ e^{-m^{2}_{e}\varrho} }{\varrho^{3}}\left[ \left(q_{e}\varrho\right)^{2}\frac{ {\tt Re}\!\left\{ \cosh\left( q_{e}\varrho \sqrt{ 2\mathcal{F}+ 2i\mathcal{G} }  \right) \right\} }{{\tt Im}\!\left\{ \cosh\left( q_{e}\varrho \sqrt{ 2\mathcal{F}+ 2i\mathcal{G} }  \right) \right\}}\mathcal{G} - \frac{2}{3}\left(q_{e}\varrho\right)^{2}\mathcal{F} - 1 \right] d\varrho
\end{equation}
where $m_{e}$ is the mass of the electron and $q_{e}$ is the elementary charge.  
Writing $\mathcal{L}_{_{\mathrm{EH}}}$ 
as a power series of $\mathcal{F}$ and $\mathcal{G}$ yields 
\begin{equation}\label{EH1}
\mathcal{L}_{_{\mathrm{EH}}} = \mathcal{L}_{_{_{\mathrm{LED}}}} - \frac{16\alpha^{2}}{45m^{4}_{e}}\left( \frac{\mathcal{F}^{2}}{2} + \frac{7}{8}\mathcal{G}^{2}\right) + \mathcal{O}\!\left(\mathcal{F}^{3},\mathcal{G}^{3},\mathcal{F}^{2}\mathcal{G},\mathcal{F}\mathcal{G}^{2}\right)
\end{equation}
where $\alpha=q^{2}_{e}/(4\pi)$ is the fine structure constant.
Assuming that the electromagnetic field is sufficiently small, and taking into account the terms up to the quadratic order of $\mathcal{F}$ and $\mathcal{G}$, the EH Lagrangian is approximated
by
\begin{equation}\label{EH2}
\mathcal{L}_{_{\mathrm{EH}}} = \mathcal{L}_{_{_{\mathrm{LED}}}} +  \frac{\gamma}{2}\mathcal{F}^{2} + \frac{7\gamma}{8}\mathcal{G}^{2} 
\end{equation}
which corresponds to the weak field approximation of \cite{Heuler_35,Heisenberg_36},  and the coupling constant $\gamma$ is written as $\gamma = - \frac{16\alpha^{2}}{45m^{4}_{e}}$. 
In addition to Born-Infeld and Euler-Heisenberg theories, other types of  
$\mathcal{L}(\mathcal{F}, \mathcal{G})$ electrodynamics
(containing such instances as power law \cite{Hassaine08,Gurtug12}, inverse \cite{Cembranos15,Gaete21}, exponential \cite{ExpNLED}, rational \cite{RationalNLED}, logarithmic \cite{LogNLED}, double logarithmic \cite{2LogNLED} and other NLEDs) have been discussed in the literature.
They  have been created for applications in gravity and cosmology, as well as for a gravity/CMT holographic description of certain strongly coupled condensed matter systems \cite{Sorokin21}.

The main objective of this article is to  present a method to generate magnetically charged ultrastatic 
nonexotic traversable wormhole solutions in Einstein-scalar-Gauss-Bonnet gravity coupled to $\mathcal{L}(\mathcal{F})$ electrodynamics (depending only on the invariant $\mathcal{F}$). 
In our case the $\boldsymbol{f}$GB curvature  is what creates the necessary violation of the null-energy condition required for the traversability of wormholes, thus, in a sense, the effective negative energy density comes from the geometry itself instead of the matter source as in GR. 
In the next section we obtain the field equations for the EsGB-$\mathcal{L}(\mathcal{F})$ gravity.  
Then in Sec. \ref{III} the generic metric of a static, spherically symmetric and asymptotically flat traversable wormhole spacetime is presented, and we examine the relation between traversability and
the requirement of exotic matter in GR and EsGB gravity contexts. In \ref{Metodo} a simple method to generate ultrastatic spherically symmetric and asymptotically flat T-WH solutions without exotic matter in EsGB-$\mathcal{L}(\mathcal{F})$ is presented, and some examples of T-WHs supported by purely magnetic fields and $\boldsymbol{f}$GB curvature
are displayed in \ref{Ejemplos}. 
This is followed by the conclusion and discussion. In this paper we use units where $G = c = \hbar =\epsilon_{_{0}}=\mu_{_{0}} =1$, and the metric signature $(-+++)$. Greek indices run from 0 to 3 and Latin indices run from 1 to 3.

\section{Basic field equations in Einstein-scalar-Gauss-Bonnet-$\boldsymbol{\mathcal{L}(\mathcal{F})}$ gravity }

In this section we shall briefly describe the dynamical equations of Einstein-scalar-Gauss-Bonnet theory 
minimally coupled to linear/nonlinear electrodynamics $\mathcal{L}(\mathcal{F})$ acting as a source.
The general action for an EsGB-$\mathcal{L}(\mathcal{F})$ theory of gravity is given by
\begin{equation}\label{actionL}
S[g_{\mu\nu},\phi,A_{\alpha}] = \int d^{4}x \sqrt{-g} \left\{ \frac{1}{16\pi}\left(R - \frac{1}{2}\partial_{\alpha}\phi\partial^{\alpha}\phi + \boldsymbol{f}(\phi) R_{_{GB}}^{2} \right) - \frac{1}{4\pi}\mathcal{L}(\mathcal{F}) \right\}
\end{equation}
which is basically the action (\ref{actionLEsGB}) with $\mathcal{L}_{\rm matter} = - \mathcal{L}(\mathcal{F})/(4\pi)$, where  $\mathcal{L}(\mathcal{F})$ is a function of the electromagnetic invariant $\mathcal{F}\equiv \frac{1}{4}F_{\alpha\beta}F^{\alpha\beta}$,
where $F_{\alpha\beta}=2\partial_{[\alpha}A_{\beta]}$ are the components of the electromagnetic field tensor $\boldsymbol{F}=\frac{1}{2}F_{\alpha\beta} \boldsymbol{dx^{\alpha}} \wedge \boldsymbol{dx^{\beta}}$, and $A_{\alpha}$ the components of the electromagnetic potential.
\\[2mm]
Using the notation $\mathcal{L}_{_{\mathcal{F}}}\equiv \frac{d\mathcal{L}}{d\mathcal{F}}$ and $\dot{\boldsymbol{f}} = \frac{d\boldsymbol{f}}{d\phi}$ (we shall use similar notation for higher derivatives), 
the EsGB-$\mathcal{L}(\mathcal{F})$ field equations arising from 
varying the action (\ref{actionL}) with respect to the metric tensor $g_{\mu\nu}$, the electromagnetic potential $A_{\alpha}$ and the scalar field $\phi$, are given by 
\begin{eqnarray}
&& G_{\alpha}{}^{\beta} + \Theta_{\alpha}{}^{\beta} = 8\pi (E_{\alpha}{}^{\beta})\!_{_{_{S\!F}}} + 8\pi T_{\alpha}{}^{\beta} \label{ESGB_NLED_Eqs} \\ 
&& \nabla_{\mu}(\mathcal{L}_{_{\mathcal{F}}}F^{\mu\nu}) = 0 = d\boldsymbol{F}
\label{NLED_Eqs} \\ 
&& \nabla^{2}\phi + \dot{\boldsymbol{f}}(\phi) R_{_{GB}}^{2} = 0,
\label{SF_Eq} %
\end{eqnarray}  
where $G_{\alpha}{}^{\beta} = R_{\alpha}{}^{\beta} - \frac{R}{2} \delta_{\alpha}{}^{\beta}$  are the components of the Einstein tensor, whereas the quantities $\Theta_{\alpha}{}^{\beta}$, $(E_{\alpha}{}^{\beta})\!_{_{_{S\!F}}}$ and $T_{\alpha}{}^{\beta}$ are defined by
\begin{eqnarray}
&& \Theta_{\alpha}{}^{\beta} 
= \frac{1}{2}( g_{\alpha\rho} \delta_{\lambda}{}^{\beta} + g_{\alpha\lambda} \delta_{\rho}{}^{\beta})\eta^{\mu\lambda\nu\sigma}\tilde{R}^{\rho\xi}{}_{\nu\sigma}\nabla_{\xi}\partial_{\mu}\boldsymbol{f}(\phi)\label{E_GB}\\
&& 8\pi (E_{\alpha}{}^{\beta})\!_{_{_{S\!F}}} 
= -\frac{1}{4}(\partial_{\mu}\phi\partial^{\mu}\phi)\delta_{\alpha}{}^{\beta} + \frac{1}{2}\partial_{\alpha}\phi \partial^{\beta}\phi  \label{SF_T}\\ 
&& 8\pi T_{\alpha}{}^{\beta}  
= 2\mathcal{L}_{_{\mathcal{F}}}F_{\alpha\mu}F^{\beta\mu} - 2\mathcal{L}\hskip.06cm\delta_{\alpha}{}^{\beta} \label{NLED_EM}
\end{eqnarray}
with $\tilde{R}^{\rho\gamma}{}_{\mu\nu} = \eta^{\rho\gamma\sigma\tau}R_{\sigma\tau\mu\nu} = \epsilon^{\rho\gamma\sigma\tau}R_{\sigma\tau\mu\nu}/\sqrt{-g}$. Thus, the quantities  $\Theta_{\alpha}{}^{\beta}$ are the components of a tensor which we refer to as the $\boldsymbol{f}$GB curvature tensor, since that captures the contribution to the spacetime curvature due to the effects of the $\boldsymbol{f}$GB 
term in the action; $(E_{\alpha}{}^{\beta})\!_{_{_{S\!F}}}$ are the components of the canonical energy-momentum tensor of the massless scalar field $\phi$, while $T_{\alpha}{}^{\beta}$ are the components of the energy-momentum tensor associated with the $\mathcal{L}(\mathcal{F})$ electrodynamics. 

The structure of the field equation \eqref{ESGB_NLED_Eqs}
motives the definition of the effective energy-momentum tensor, $\mathscr{E}_{\alpha}{}^{\beta}$, as $8\pi\mathscr{E}_{\alpha}{}^{\beta} = 8\pi(E_{\alpha}{}^{\beta})\!_{_{_{S\!F}}} - \Theta_{\alpha}{}^{\beta} +  8\pi T_{\alpha}{}^{\beta}$, thus, the  field equation \eqref{ESGB_NLED_Eqs}
can be written in a GR-like form as $G_{\alpha}{}^{\beta} = 8\pi \mathscr{E}_{\alpha}{}^{\beta}$. 

Now, we focus on obtaining the relevant motion equations of the EsGB-$\mathcal{L}(\mathcal{F})$ system (without cosmological constant)
for a static and spherically symmetric spacetime  
with a magnetic field as the source.  
To do this, we assume  
the static and spherically symmetric spacetime metric in standard coordinates   
$(t,r,\theta,\varphi)$, also known as
Schwarzschild like coordinates, is given by
\begin{equation}\label{SSSmet}
ds^{2} =  - e^{ A(r) }dt^{2} + e^{ B(r) }dr^{2}  + r^{2}(d\theta^{2}  + \sin^{2}\theta d\varphi^{2})  
\end{equation}
with  $A(r)$ and $B(r)$ unknown functions to be determined, and also the scalar field is static and spherically symmetric, $\phi = \phi(r)$. 
Taking into account the line element (\ref{SSSmet}), the nonvanishing components of the Einstein tensor can be expressed as
\begin{equation}\label{GabSSS}
G_{t}{}^{t}\!=\!\frac{ e^{^{\!\!-B}} }{ r^{2} }\!\!\left( -rB' - e^{^{\!B}} + 1 \right)\!, \hspace{0.25cm}  G_{r}{}^{r} \!=\! \frac{ e^{^{\!\! -B}} }{ r^{2} }\!\!\left( rA' - e^{^{\!B}} + 1 \right)\!, \hspace{0.25cm} G_{\theta}{}^{\theta}\!=\! G_{\varphi}{}^{\varphi}\!=\!\frac{ e^{^{\!\!-B}} }{ 4r }\!\!\left( rA'^{2} - rA'B' + 2rA'' + 2A' - 2B' \right),
\end{equation}
where the prime $'$ denotes the  derivative with respect to the radial coordinate $r$ ({\it i.e.} $A' = \frac{dA}{dr}$).
For the nontrivial components of the energy-momentum tensor of self-interacting scalar field we have
\begin{equation}\label{EttyErr}
8\pi (E_{t}{}^{t})_{\!_{SF}} = 8\pi (E_{\theta}{}^{\theta})_{\!_{SF}} = 8\pi (E_{\varphi}{}^{\varphi})_{\!_{SF}} = - 8\pi (E_{r}{}^{r})_{\!_{SF}} = -\frac{1}{ 4 } e^{ -B} \phi'^{2}.  
\end{equation}
The non-vanishing components of the $\boldsymbol{f}$GB 
curvature tensor are
\begin{eqnarray}\label{einstensors}
&& \Theta_{t}{}^{t} = \frac{ e^{ -2B} }{ 4r^{2} }\left\{  16(e^{B} - 1)\ddot{\boldsymbol{f}} \phi'^{2} - 8[ (e^{B} - 3)B'\phi' - 2(e^{B} - 1)\phi'']\dot{\boldsymbol{f}} \right\}  
\label{GBtt} \\
&& \Theta_{r}{}^{r} =  \frac{8(e^{B} - 3)e^{-2B}A' \phi' \dot{\boldsymbol{f}} }{4r^{2}}   
\label{GBrr} \\
&& \Theta_{\theta}{}^{\theta} = \Theta_{\varphi}{}^{\varphi} = -\frac{ e^{ -2B} }{ 4r } \left\{   8A'\ddot{\boldsymbol{f}}\phi'^{2} + 4\left[ (A'^{2} + 2A'')\phi' + (2\phi'' - 3B'\phi')A'\right]\dot{\boldsymbol{f}} \right\}. 
\label{einstensors2}
\end{eqnarray}

Regarding the electromagnetic field tensor, since the spacetime is static and spherically symmetric, then the only nonvanishing terms are the electric component $F_{tr}=\mathcal{E}(r)$ and the magnetic component $F_{\theta\varphi}=\mathcal{B}(r,\theta)$. However, in this work, we can restrict ourselves to the purely magnetic field, {\it i.e.} $F_{tr} = 0$ and $F_{\theta\varphi} \neq 0$.
With this restriction, the electromagnetic field tensor has the form $F_{\alpha\beta} = \left( \delta^{\theta}_{\alpha}\delta^{\varphi}_{\beta} - \delta^{\varphi}_{\alpha}\delta^{\theta}_{\beta} \right)F_{\theta\varphi}$.  
In this way, for a static and spherically symmetric spacetime with line element (\ref{SSSmet}), the general solution of the equations $\nabla_{\mu}(\mathcal{L}_{_{\mathcal{F}}}F^{\mu\nu})=0$ is
\begin{equation}\label{fabSOL}
F_{\theta\varphi} =  r^{4} \mathrm{Q}(r) \sin\theta,
\end{equation}
where $\mathrm{Q}(r)$ is a function of $r$ only. 
This means  $\boldsymbol{F} = r^{4} \mathrm{Q}(r)\sin\theta \hspace{0.1cm} \boldsymbol{d\theta} \wedge \boldsymbol{d\varphi}$, and therefore $d\boldsymbol{F} = \left(r^{4} \mathrm{Q}(r)\right)' \sin\theta \hspace{0.1cm} \boldsymbol{dr} \wedge \boldsymbol{d\theta} \wedge \boldsymbol{d\varphi} = 0$. This implies \begin{math}\mathrm{Q}(r)=\sqrt{2}\hskip.06cm q/r^{4}\end{math}, where
$\sqrt{2}\hskip.06cm q$ is an integration constant  associated with  
magnetic charge. 
Hence, in the pure magnetic sector, the components of the electromagnetic field tensor $F_{\alpha \beta}$ and the invariant $\mathcal{F}$ are
given by
\begin{equation}\label{magnetica}
F_{\alpha\beta} = \sqrt{2} \hspace{0.05cm} q  \sin \theta \hspace{0.05cm} \left( \delta^{\theta}_{\alpha}\delta^{\varphi}_{\beta} - \delta^{\varphi}_{\alpha}\delta^{\theta}_{\beta} \right),  \quad\quad\quad\quad \mathcal{F} =  \frac{q^{2}}{r^{4}}.
\end{equation}
Finally, the nonvanishing components of $T_{\alpha}{}^{\beta}$ for an arbitrary $\mathcal{L}(\mathcal{F})$ function, considering the standard static and spherically symmetric line element ansatz (\ref{SSSmet}) and a purely magnetic field  (\ref{magnetica}), are written as
\begin{eqnarray}\label{E_nled}
8\pi T_{t}{}^{t} = 8\pi T_{r}{}^{r} =  -2\mathcal{L},   
\quad\quad\quad 8\pi T_{\theta}{}^{\theta} = 8\pi T_{\varphi}{}^{\varphi} =  2(2\mathcal{F}\mathcal{L}_{_{\mathcal{F}}} - \mathcal{L}). 
\end{eqnarray}
%
After replacing the components  (\ref{GabSSS})$-$(\ref{einstensors2}) and (\ref{E_nled}), in the field equation (\ref{ESGB_NLED_Eqs}), we obtain
%
\begin{eqnarray}
&&\!G_{t}{}^{t}\!=\!8\pi\mathscr{E}_{t}{}^{t}\!\hskip.2cm\Rightarrow\hskip.2cm 
4e^{B}\!\!\left( rB' \!+\! e^{ B} \!-\! 1 \right) \!=\!\!\left[ r^{2}e^{B}\!+\!16(e^{ B}\!-\!1)\ddot{\boldsymbol{f}} \right]\!\!\phi'^{2} \!-\!8\!\left[ (e^{ B} \!-\!3)B'\phi' \!-\! 2(e^{ B} \!-\! 1)\phi'' \right]\!\!\dot{\boldsymbol{f}} 
\!+\! 8r^{2}e^{2B}\mathcal{L} \label{Eqt}\\
&&\nonumber\\   
&&\!G_{r}{}^{r}\!=\!8\pi\mathscr{E}_{r}{}^{r}\!\hskip.115cm\Rightarrow\hskip.115cm 4e^{B}\!\!\left( -rA'\!+\! e^{ B} \!-\! 1 \right) \!=\! - r^{2}e^{B}\phi'^{2} \!+\! 8(e^{ B} \!-\! 3)A'\phi'\dot{\boldsymbol{f}} \!+\! 8r^{2}e^{2B}\mathcal{L}
\label{Eqr}\\
&&\nonumber\\
&&\!G_{\theta}{}^{\theta}\!=\!8\pi\mathscr{E}_{\theta}{}^{\theta}\!\hskip.115cm\Rightarrow\hskip.115cm e^{B}\!\!\left[
rA'^{2} \!-\! 2B' \!+\! (2 \!-\! rB')A' \!+\! 2rA'' \right] \!=\! -re^{B}\phi'^{2} \!+\! 8A'\ddot{\boldsymbol{f}}\phi'^{2} \nonumber \\
&& \hskip6.9cm \!+ 4\!\left[ (A'^{2} \!+\! 2A'')\phi' \!+\! (2\phi'' \!-\! 3B'\phi')A' \right]\!\!\dot{\boldsymbol{f}} \!-\! 8 r e^{2B}(\mathcal{L}\!-\!2\mathcal{F}\mathcal{L}_{_{\mathcal{F}}}).\label{Eqte}
\end{eqnarray}
The equation of motion for the scalar field, Eq. \eqref{SF_Eq}, can be written as
\begin{equation}\label{phi2} 
2r\phi'' + (4 + rA' - rB')\phi' + \frac{4e^{-B}\dot{\boldsymbol{f}}}{r} \left[ (e^{B} - 3)A'B' - (e^{B} - 1)(2A'' + A'^{2})\right] = 0.
\end{equation}  
This ends the general treatment of the static, spherically symmetric and pure magnetic solutions in EsGB-$\mathcal{L}(\mathcal{F})$. 
\section{The generic metric of a static, spherically symmetric and asymptotically flat traversable wormhole spacetime }\label{III} 
In accordance with \cite{morris88,morris88-2}, the generic 
metric of a (3+1)-dimensional static, spherically symmetric and asymptotically flat traversable wormhole spacetime of the Morris-Thorne type is given by
\begin{equation}\label{TWH_MT}
\boldsymbol{ds^{2}} = - e^{2\Phi(r)}\boldsymbol{dt^{2}} + \frac{\boldsymbol{dr^{2}}}{ 1 - \frac{b(r)}{r} } + r^{2}\boldsymbol{d\Omega^{2}},
\end{equation}  
where $\Phi(r)$ and $b(r)$ are smooth functions, respectively known as redshift function (since this gives a measure of the gravitational redshift) and shape function (since this determines the topological configuration of the spacetime). 
Moreover, the WH spacetime is characterized by the existence of a throat, where this is a two-sphere of radius $r_{_{0}}$ satisfying $b(r_{_{0}}) = r_{_{0}}$, acting as a membrane connecting the two sides of the WH.
The range of the $r$ coordinate, is $r\in[r_{_{0}},\infty)$. Therefore, here the coordinate $r$ has a special geometric interpretation, such that  $4\pi r^2$ is the area of a sphere centered on the WH throat.
On the other hand, as was shown in Refs. \cite{morris88,morris88-2}, for the WH to be traversable, the fulfilling of the following conditions is required: 
\begin{eqnarray} 
&&\textup{{\it Wormhole domain:}}\hspace{2.3cm}   
 1 - \frac{b(r)}{r} > 0  \quad\quad\quad\quad \forall\quad\!\!\!\!\!r>r_{_{0}} \label{TWC1}\\[2.7pt]
&& \textup{{\it Absence of horizons:}}\hspace{2cm}  e^{2\Phi(r)}\in\mathbb{R}^{+}\!-\!\{0\}  
\quad\quad  \forall\quad\!\!\!\!\! r\geq r_{_{0}} \label{TWC2}\\[2.7pt]
&&\textup{{\it Flare-out condition:}}\hspace{2.1cm} b'(r)\Big|_{r=r_{_{0}}}<1\label{TWC3}\\
&&\textup{{\it AF spacetime:}}\hspace{3.0cm} \lim\limits_{r \to \infty}\Phi(r)=0 \quad\quad\textup{and} 
\quad\quad \lim\limits_{r \to \infty} \frac{b(r)}{r}=0 \label{AFWH}
\end{eqnarray}
where, as well as in Eq.\eqref{GabSSS}, the prime $'$ denotes derivative with respect to $r$. 

Let us identify the metrics  (\ref{SSSmet}) and (\ref{TWH_MT}), that is 
$e^{A(r)} ~=~e^{2\Phi(r)}$  and $e^{B(r)} = \left(1 - \frac{b(r)}{r} \right)^{^{\!-1}}$ and consider the null vector $\boldsymbol{n}=e^{-\Phi(r)}\boldsymbol{\partial_{t}} + \left( 1  - \frac{b(r)}{r} \right)^{\frac{1}{2}}\boldsymbol{\partial_{r}}$, written as a linear combination of elements of a basis $\{ \boldsymbol{\partial_{t}}, \boldsymbol{\partial_{r}}, \boldsymbol{\partial_{\theta}}, \boldsymbol{\partial_{\varphi}} \}$ of vector space, whereas $\{ \boldsymbol{dt}, \boldsymbol{dr}, \boldsymbol{d\theta}, \boldsymbol{d\varphi} \}$ is a basis of dual vector space, such that $dx^{\alpha}(\partial_{\beta}) = \partial_{\beta}(dx^{\alpha}) = \delta_{\beta}^{\alpha}$. We can see that  $\boldsymbol{n}$ is a null vector since $ds^{2}( \boldsymbol{n}, \boldsymbol{n}) = g_{_{\alpha\beta}}n^{\alpha}n^{\beta} = 0$. 
Using  \eqref{GabSSS} and assuming  the flaring out condition is fulfilled [{\it i.e.} $b'(r_{_{0}})<1$],  after contracting the Einstein tensor with $\boldsymbol{n}$ and evaluating at $r=r_{_{0}}$, yields
\begin{eqnarray}\label{Gnnr0}
 G_{\alpha\beta}n^{\alpha}n^{\beta}  \Big|_{r=r_{_{0}}} = \left(G_{r}{}^{r} - G_{t}{}^{t}\right)\Big|_{r=r_{_{0}}} 
 =\frac{1}{r^{2}_{_{0}}} \left[ b'(r_{_{0}})- 1 \right]< 0. 
\end{eqnarray}
Below, let us examine  the effects of inequality \eqref{Gnnr0} in the gravity contexts of GR and EsGB-$\mathcal{L}(\mathcal{F})$.

\begin{itemize}

 \item {\bf Traversability of wormhole and violation of Null Energy Condition (NEC) in GR.} \\\noindent
Let us consider the construction of a T-WH solution (\ref{TWH_MT}), in the context of
GR.
Then, demanding the fulfillment of the Einstein field equations $G_{\alpha\beta} = 8\pi T_{\alpha\beta}$, the condition (\ref{Gnnr0}) implies that
\begin{equation}
T_{\alpha\beta}n^{\alpha}n^{\beta}\!\Big|_{r=r_{0}} < 0.
\end{equation}
This yields that in GR the NEC (which establishes that for any null vector $n^{\alpha}$, $T_{\alpha\beta}n^{\alpha}n^{\beta}\geq0$) is violated for a T-WH spacetime.
Therefore, in GR the fulfillment of the flaring out condition, which is considered fundamental for a T-WH, implies the existence of exotic matter ({\it i.e.} matter whose energy momentum tensor violates the NEC).
In summary, in the GR context, the violations of the energy conditions at wormhole throats
are unavoidable.

 \item {\bf Traversability of wormhole and fulfillment of NEC in EsGB-$\boldsymbol{ \mathcal{L}(\mathcal{F}) } $.} \\\noindent
Now, with regards to EsGB-$\mathcal{L}(\mathcal{F})$, the field equation \eqref{ESGB_NLED_Eqs} has an analogous structure to those in GR with a effective energy-momentum tensor given by $8\pi\mathscr{E}_{\alpha\beta} = 8\pi(E_{\alpha\beta})\!_{_{_{S\!F}}} - \Theta_{\alpha\beta} +  8\pi T_{\alpha\beta}$.   
In the same way as \eqref{Gnnr0},  us to calculate $\mathscr{E}_{\alpha\beta}n^{\alpha}n^{\beta}$ using the null vector $\boldsymbol{n}=e^{-\Phi(r)}\boldsymbol{\partial_{t}} + \left( 1  - \frac{b(r)}{r} \right)^{\frac{1}{2}}\boldsymbol{\partial_{r}}$. Thus, one finds that
\begin{equation}
8\pi\mathscr{E}_{\alpha\beta}n^{\alpha}n^{\beta} =  8\pi\left[ (E_{r}{}^{r})\!_{_{_{S\!F}}} - (E_{t}{}^{t})\!_{_{_{S\!F}}}  \right] - \Big[ \Theta_{r}{}^{r} - \Theta_{t}{}^{t}  \Big] + 8\pi\left[ T_{r}{}^{r} - T_{t}{}^{t}  \right].
\end{equation}
From (\ref{E_nled}) it follows that $T_{r}{}^{r} - T_{t}{}^{t}  = 0$, and hence
\begin{equation}\label{E_GB_null}
8\pi \mathscr{E}_{\alpha\beta}n^{\alpha}n^{\beta} = 8\pi \left[ (E_{r}{}^{r})\!_{_{_{S\!F}}} - (E_{t}{}^{t})\!_{_{_{S\!F}}} \right] - \Big[ \Theta_{r}{}^{r} - \Theta_{t}{}^{t} \Big].
\end{equation}
Taking advantage of the structure of the EsGB field equation \eqref{ESGB_NLED_Eqs} and using (\ref{Gnnr0}), we conclude that in EsGB-$\mathcal{L}(\mathcal{F})$ theory (in the pure magnetic sector), for a traversable wormholes spacetime (\ref{TWH_MT}), the flaring out condition (\ref{TWC3}) implies that
\begin{equation}\label{E_GB_null_vio}
8\pi\mathscr{E}^{(e\!f\!f)}_{\alpha\beta}n^{\alpha}n^{\beta}\!\Big|_{r=r_{0}} = 8\pi\left[ (E_{r}{}^{r})\!_{_{_{S\!F}}} - (E_{t}{}^{t})\!_{_{_{S\!F}}} \right]\!\Big|_{r=r_{0}} - \Big[ \Theta_{r}{}^{r} - \Theta_{t}{}^{t} \Big]\!\Big|_{r=r_{0}} < 0.  
\end{equation}
But, according to \eqref{EttyErr}, if the scalar field is real defined [$\phi(r)\in \mathbb{R}$] in the whole T-WH spacetime, this yields 
\begin{equation}\label{E_GB_scalar_vio}
(E_{r}{}^{r})\!_{_{_{S\!F}}} - (E_{t}{}^{t})\!_{_{_{S\!F}}} = \frac{1}{2} \left( 1 - \frac{b(r)}{r} \right) \phi'^{2} > 0 \quad\quad\quad\quad \forall~ r \geq r_{_{0}}; 
\end{equation}
hence, the $\boldsymbol{f}$GB curvature is solely responsible for fulfilling the inequality \eqref{E_GB_null_vio}.
Therefore, in the framework of EsGB-$\mathcal{L}(\mathcal{F})$ gravity, the existence of nonexotic static and spherically symmetric traversable wormhole solutions is not ruled out. 
Now we are ready to present a general procedure to derive ultrastatic T-WH spacetime solutions supported by
nonexotic matter in this modified gravity context.
\end{itemize}
\section{A simple method to generate  ultrastatic spherically symmetric and asymptotically flat traversable wormholes 
without exotic matter in $\boldsymbol{ \textup{EsGB}}$-$\boldsymbol{\mathcal{L}(\mathcal{F})}$ }\label{Metodo}

We start by mentioning some general aspects about the ultrastatic spherically symmetric and asymptotically flat spacetimes.
A spacetime is called ultrastatic if it admits an atlas of charts in which the metric tensor takes the
form
\begin{equation}\label{UltraS}
\boldsymbol{ds^{2}} = - \boldsymbol{dt^{2}} + g_{jl}\boldsymbol{dx^{j}}\boldsymbol{dx^{l}}
\end{equation}  
in some coordinate system $\{ x^{\alpha} \}^{^{3}}_{_{\alpha=0}} = \left( t,\{x^{l}\}^{^{3}}_{_{l=1}} \right)$, where the Latin indices label spatial coordinates only, 
where $t$ is the time relative to a free-falling observer moving with four-velocity $u^{\alpha} = - g^{\alpha\beta}\partial_{\beta}t$, 
and where the metric coefficients $g_{jl}$ are independent of the time coordinate $t$. 
In an ultrastatic spacetime the only nonvanishing Christoffel symbols are
$\Gamma^{i}{}_{jl}$ (and their partial derivatives with respect to spatial coordinates), see \cite{Stephani1900} for details. 
In other words, computing the Christoffel symbols for the metric \eqref{UltraS}, we get
\begin{equation}
\Gamma^{0}{}_{\alpha\beta} = \Gamma^{\alpha}{}_{\beta0} = \partial_{_{0}}\Gamma^{i}{}_{jl} = 0      
\end{equation}
which implies that the only components of the Riemann tensor $R^{\alpha}{}_{\beta\mu\nu}$, of the metric (\ref{UltraS}), that do not vanish identically, coincide with those of the Riemann tensor $^{^{(3)}}\!\!R^{\alpha}{}_{\beta\mu\nu}$ of the three-dimensional metric $\boldsymbol{^{^{(3)}}\!\!ds^{2}}=g_{jl}\boldsymbol{dx^{j}}\boldsymbol{dx^{l}}$.  As a consequence, the differential geometrical properties of ultrastatic spacetime \eqref{UltraS} are completely determined by  the Riemannian metric induced on three-dimensional hypersurface $t$ = constant, 
$\boldsymbol{^{^{(3)}}\!\!ds^{2}}=g_{jl}\boldsymbol{dx^{j}}\boldsymbol{dx^{l}}$ (see  \cite{Sonego2010,Fulling,Fulling81,DonPage} for details).  

For the particular case of interest, setting $\Phi(r)=0$ in the line element ansatz \eqref{TWH_MT} and preserving the conditions \eqref{TWC1}$-$\eqref{AFWH}, one arrive to the canonical metric for an ultrastatic, spherical symmetric and asymptotically flat T-WH spacetime. 

\subsection{\bf 
Method to generate ultrastatic spherically symmetric spacetime solutions in 
EsGB-$\boldsymbol{\mathcal{L}(\mathcal{F})}$ } 
Below, the  more general magnetically charged ultrastatic and spherically symmetric spacetime solution  of  EsGB-$\mathcal{L}(\mathcal{F})$ gravity is presented.
%

{\it Method:} starting from an arbitrary $\mathcal{L}(\mathcal{F})$ model, the following shape function 
\begin{equation}\label{Fsol}   
b(r) = \frac{8r^{4}\mathcal{L}(\mathcal{F}) - \sigma^{2}_{_{0}} }{4r} \quad\quad \textup{with} \quad\quad \mathcal{F} =  \frac{q^{2}}{r^{4}},  
\end{equation}
where $\sigma_{_{0}}$ and $q$ are real parameters (representing the scalar and magnetic charge, respectively), determines a magnetically charge ultrastatic and spherically symmetric spacetime metrics given by
\begin{equation}\label{UltraS_Solution} 
ds^{2} =  - dt^{2} + \frac{dr^{2}}{ 1 - \frac{b(r)}{r} }  + r^{2}(d\theta^{2}  + \sin^{2}\theta d\varphi^{2})
\end{equation}
which is an exact solution of EsGB-$\mathcal{L}(\mathcal{F})$ 
field equations \eqref{Eqt}$-$\eqref{phi2}, with EsGB theory  determined by a massless scalar field and coupling function (in terms of the radial coordinate), given respectively by 
\begin{equation}\label{phi_sol}
\phi(r) = \int^{r}_{r_{_{0}}} \frac{\sigma_{_{0}}}{\chi^{2}} \left( 1 - \frac{b(\chi)}{\chi}\right)^{-\frac{1}{2}}d\chi, \quad\quad\quad\quad\quad \boldsymbol{f}(r) =\int^{r}_{r_{_{0}}} \phi'(\tilde{r})\dot{\boldsymbol{f}}(\tilde{r})d\tilde{r},
\end{equation}
with $\dot{\boldsymbol{f}}(r)$ given by 
\begin{equation}\label{df_sol}
\dot{\boldsymbol{f}}(r) = \frac{r^{3}}{ \sigma_{_{0}} b(r) } \int^{r}_{r_{_{0}}} \frac{ \left( 2\chi^{2} b'(\chi) - 2\chi b(\chi) - \sigma^{2}_{_{0}} \right) }{8\chi^{2}} \left( 1 - \frac{b(\chi)}{\chi}\right)^{-\frac{1}{2}}d\chi,         
\end{equation}   
where $r_{_{0}}$ is a integration constant, which we are setting as the minimum value of the radial coordinate $r$, {\it i.e.} $r\in [r_{_{0}} , \infty )$. Therefore, given a linear/nonlinear electrodynamics with an arbitrary electromagnetic Lagrangian density $\mathcal{L}(\mathcal{F})$, being $\mathcal{L}(\mathcal{F})$ a well-definite function of $\mathcal{F}$, it follows that the metric \eqref{UltraS_Solution} with shape function \eqref{Fsol}, together with the scalar field \eqref{phi_sol} and sGB coupling function \eqref{df_sol}, is the more general purely magnetic ultrastatic and spherically symmetric solution of EsGB-$\mathcal{L}(\mathcal{F})$ field equations \eqref{Eqt}$-$\eqref{phi2}. 

It is important to stress that Eqs. \eqref{Fsol}$-$\eqref{df_sol} come from integrating the field equations \eqref{Eqt}$-$\eqref{phi2}, for an arbitrary magnetically charged ultrastatic spherically symmetric spacetime geometry. 
This means that any magnetically charged ultrastatic and spherically symmetric EsGB-$\mathcal{L}(\mathcal{F})$ solution must be compatible with Eqs. \eqref{Fsol}$-$\eqref{df_sol}.

We now turn to the form of the solution of the field equations at large distances from
the spacetime region $r = r_{_{0}}$. 
Although Eqs. \eqref{Fsol}$-$\eqref{df_sol} are valid for any 
$\mathcal{L}(\mathcal{F})$ model, not all
those models correspond to physically reasonable models of 
electrodynamics. In order to do this, several assumptions 
on the Lagrangian density 
$\mathcal{L}(\mathcal{F})$ must be considered. 
An important assumption is to require that in the weak field limit ($\mathcal{F} \rightarrow 0 \equiv r \rightarrow \infty$) the Lagrangian density $\mathcal{L}(\mathcal{F})$ approaches the Maxwell Lagrangian $\mathcal{L}(\mathcal{F}) \approx \mathcal{F}$. Thus, for a $\mathcal{L}(\mathcal{F})$ model satisfying the weak field limit yields that at the asymptotic region ($r \rightarrow \infty $) the shape function \eqref{Fsol} goes to zero as $b(r) \approx (8q^{2} - \sigma^{2}_{_{0}})/(4r)$,  and the scalar field \eqref{phi_sol} has the following asymptotic behavior:
\begin{equation}
\phi(r) = \phi_{_{\infty}} - \frac{\sigma_{_{0}}}{r} + \mathcal{O}\!\!\left(\!\frac{1}{r^{2}}\!\right).
\end{equation}
Thus, the parameter $\sigma_{_{0}}$  can be associated  with the charge of the scalar field \eqref{phi_sol}, and $\phi_{_{\infty}}$ is a parameter  that denotes the asymptotic value of the scalar field\footnote{ That is, $\lim\limits_{r \to \infty} \phi(r) = \lim\limits_{r \to \infty} \int^{r}_{r_{_{0}}} \frac{\sigma_{_{0}}}{\chi^{2}} \!\!\left( 1 - \frac{b(\chi)}{\chi}\right)^{\!\!\!^{-\frac{1}{2}}}d\chi = \phi_{_{\infty}}$ for the $\mathcal{L}(\mathcal{F})$ model with a Maxwell asymptotic limit.  } (see \cite{antoniou19} for details). 

Notice that a magnetically charged ultrastatic and spherically symmetric EsGB-$\mathcal{L}(\mathcal{F})$ solution,
with $\mathcal{F}(r)=\frac{q^{2}}{r^{4}}$ 
having a strict global maximum\footnote{ That is, $\mathcal{F}_{_{0}}=\mathcal{F}(r_{_{0}}) > \mathcal{F}(r)$ for all $r\neq r_{_{0}}$ in the entire range of the radial coordinate. } $\mathcal{F}_{_{0}}$ at $r_{_{0}}$, 
and such that  
$\mathcal{L}_{_{0}} = \mathcal{L}(\mathcal{F}_{_{0}})$ is a positive number ($\mathcal{L}_{_{0}}>0$),
admits
a magnetically charged ultrastatic wormhole interpretation 
with throat radius given by  
\begin{equation}
 r_{_{0}} = \sqrt{ \frac{1 + \sqrt{ 2\sigma^{2}_{_{0}}\mathcal{L}_{_{0}} + 1} }{4\mathcal{L}_{_{0}} }  }    
\end{equation}

{\it Theorem 1.} {{\it In an Einstein-scalar-Gauss-Bonnet theory whose action is given by Eq. \eqref{actionL}, where $\boldsymbol{f}(\phi)$ is a continuous
and differentiable function of the class $\mathcal{C}^{2}$ (as a minimum),  
the only ultrastatic spherically symmetric and asymptotically flat traversable wormhole solution of the pure EsGB field equations [{\it i.e.} field equations \eqref{ESGB_NLED_Eqs}$-$\eqref{SF_Eq},  
with $\mathcal{L}(\mathcal{F}) = 0$]  
is the Ellis-Bronnikov wormhole metric characterized by an imaginary scalar field  (with negative kinetic energy term) and  $\boldsymbol{f}(\phi) =$ constant, everywhere of the T-WH spacetime. }}

{\it Proof.}
The problem of describing an ultrastatic spherically symmetric and asymptotically flat T-WH solution in pure EsGB theory ({\it i.e.} in  Einstein-scalar-Gauss-Bonnet without additional matters fields)  reduces to
solving the 
equations 
\eqref{Fsol}$-$\eqref{df_sol} with $\mathcal{L}(\mathcal{F})=0$. 
For this case the shape function \eqref{Fsol} becomes $b(r) = - \sigma^{2}_{_{0}}/(4r)$;  hence, the line element \eqref{UltraS_Solution} with $\Phi(r)=0$ takes the form
\begin{equation}\label{Lfcero_metric} 
ds^{2} =  - dt^{2} + \frac{dr^{2}}{ 1 + \frac{\sigma^{2}_{_{0}} }{4r^{2}} }  + r^{2}(d\theta^{2}  + \sin^{2}\theta d\varphi^{2}).
\end{equation}

The scalar field \eqref{phi_sol}  becomes
\begin{equation}\label{Lfcero_sf}
\phi = \frac{2\sigma_{_{0}}}{\sqrt{ - \sigma^{2}_{_{0}} }}\tan^{-1}\left(  \frac{ \sqrt{ 4r^{2} + \sigma^{2}_{_{0}}  } }{ \sqrt{ - \sigma^{2}_{_{0}}} } \right) = \pm 2 \tanh^{-1}\left(   \sqrt{ \frac{ 4r^{2} + \sigma^{2}_{_{0}}  }{  \sigma^{2}_{_{0}}}  } \right).   
\end{equation}
Evaluating Eq. \eqref{df_sol} yields $\boldsymbol{f}(\phi) =$ constant. Hence, for this case, the Einstein-scalar-Gauss-Bonnet theory reduces to general relativity, and 
the spacetime metric\footnote{ In the published version of this paper, 
there is a typographical error 
in Eq. \eqref{Lfcero_metric}. } \eqref{Lfcero_metric} describes a naked singularity spacetime. 
If one allows  $\sigma_{_{0}}$ to be an imaginary number,  
that is $\sigma_{_{0}} = 2 a i$ being $a\in\mathbb{R}$ and $i^{2}=-1$, the scalar field becomes imaginary, and is given by
\begin{equation}\label{phiEllisWH}
\phi = \pm 2i\tan^{-1}\left( \sqrt{ \frac{ r^{2} - a^{2}  }{  a^{2}  } } \right)
\end{equation}
whereas the line element \eqref{Lfcero_metric} becomes
\begin{equation}\label{EllisWH}
ds^{2} = - dt^{2} + \left( 1 - \frac{a^{2}}{r^{2}}  \right)^{\!\!^{-1}}
dr^{2} + r^{2}(d\theta^{2}  + \sin^{2}\theta d\varphi^{2}). 
\end{equation}
This metric, was originally introduced in \cite{ellis73,Bronnikov73}, admits a T-WH interpretation since satisfies the properties \eqref{TWC1}-\eqref{AFWH} with wormhole throat radius $r_{_{0}}=|a|$, and is known as the Ellis-Bronnikov wormhole metric\footnote{ Transforming the standard radial coordinate $r$ into a new radial coordinate $\rho$ defined as
$\rho = \pm\sqrt{ r^{2} - a^{2} }$,  where the plus (minus) sign is related to the upper (lower) part of the wormhole, which implies $d\rho = \pm (r/\sqrt{ r^{2} - a^{2} })dr$, the line element \eqref{EllisWH} acquires a simpler structure given by
\begin{equation}\label{genr_Ellis}
ds^{2} =  - dt^{2} +  d\rho^{2}  + (\rho^{2} + a^{2} )(d\theta^{2}  + \sin^{2}\theta d\phi^{2}) 
\end{equation}
that has the form of the metric originally introduced by Ellis metric (drainhole).}.
Since for this case $\mathcal{L}(\mathcal{F})=\dot{\boldsymbol{f}}(\phi)=0$, yields that the gravitational action
(\ref{actionL}) for the which the metric (\ref{EllisWH}) together with the massless scalar field (\ref{phiEllisWH})  becomes an exact solution, takes the following form:
\begin{equation}\label{actionLphi}
S[g_{ab},\phi] = \int d^{4}x \sqrt{-g} \left\{ \frac{1}{16\pi}\left(R - \frac{1}{2}\partial_{\mu}\phi\partial^{\mu}\phi \right)  \right\}.
\end{equation}
Alternatively, defining a new scalar field by $\psi = i\phi$ (phantom field), the wormhole metric (\ref{EllisWH}) becomes a solution to the theory with gravitational action
\begin{equation}\label{actionL2}
S[g_{ab},\psi] = \int d^{4}x \sqrt{-g} \left\{ \frac{1}{16\pi}\left(R + \frac{1}{2}\partial_{\mu}\psi\partial^{\mu}\psi \right)  \right\}
\end{equation}
with $\psi$  given by
\begin{equation}\label{SfactionL2}
\psi =  2\tan^{-1}\left( \sqrt{ \frac{r^{2}  - q^{2}}{q^{2}} } \right)  \in \mathbb{R}.
\end{equation}
This is the action that was used by Ellis in Ref.\!~\cite{ellis73} to get the wormhole solution \eqref{EllisWH}, independently derived also by  Bronnikov in Ref.\!~\cite{Bronnikov73}, being this the first example of a  Morris-Thorne traversable wormhole solution in General relativity theory. 
Therefore, the Ellis-Bronnikov wormhole, whose energy momentum tensor (in the GR context) can be represented by a massless phantom scalar field with negative kinetic term plays the role of exotic matter, is considers as one of the first and simplest examples of T-WHs in GR. 
This solution has been extensively studied, and its properties like gravitational lensing \cite{EllisLensing}, quasinormal modes \cite{QNMEllis},
shadows \cite{EllisShadows} and stability \cite{EllisStability} have all been thoroughly investigated.
%
%
\section{Construction of traversable wormholes without exotic matter in Einstein-scalar-Gauss-Bonnet-$\boldsymbol{\mathcal{L}(\mathcal{F})}$ gravity 
}\label{Ejemplos}

Considering $\mathcal{L}(\mathcal{F})$ electromagnetic models coupled to GR or modified theories of gravity, interesting solutions like regular black holes and traversable wormholes, have been constructed. See, for instance, Refs. \cite{wormholes,Hassaine08,Gurtug12,Cembranos15,ExpNLED,2LogNLED,Canate2020,canate19b,canate2023,Ayon1998,Bronnikov_Ayon1998,Bronnikov2000,Canate2022}  among others. However, most NLED models used require an unreasonable amount of fine-tuning and, furthermore, lack a fundamental theoretical origin. 

Below, let us display some examples of applications of the method presented in Sec. \ref{Metodo}. 
It is important to remark that all of the $\mathcal{L}(\mathcal{F})$ models that will be used to produce our solutions have a fundamental theoretical origin. Moreover, these models in the weak field limit become the Maxwell electrodynamics, and have been used in other issues, such as the problem of infinite energy of electron as point charge, the black hole singularity problem, and cosmology (generate primordial magnetic
fields in the Universe).  

\subsection{ Magnetically charged Ellis-Bronnikov wormhole in EsGB Maxwell gravity} 
Let us consider the Maxwell's electromagnetic
theory which is defined by a linear Lagrangian in the electromagnetic invariant $\mathcal{F}$, {\it i.e.}
$\mathcal{L}_{_{_{\mathrm{LED}}}}= \mathcal{F}$. 
For this case, computing the shape function  
\eqref{Fsol} and substituting in \eqref{UltraS_Solution} yields the following ultrastatic spacetime metric:
\begin{equation}\label{EsGB_maxwell}
ds^{2} =  - dt^{2} + \left( 1 - \frac{8q^{2}-\sigma^{2}_{_{0}}}{4r^{2}}  \right)^{\!\!^{-1}}
dr^{2} + r^{2}(d\theta^{2}  + \sin^{2}\theta d\varphi^{2})
\end{equation}
whereas, according to Eqs. \eqref{phi_sol} and \eqref{df_sol}, the corresponding EsGB model for which this metric 
is an exact purely magnetic solution of the EsGB Maxwell field equations can be determined by
\begin{eqnarray}
\phi &=& \frac{2\sigma_{_{0}}}{\sqrt{ 8q^{2} - \sigma^{2}_{_{0}} }}\tan^{-1}\left(  \frac{ \sqrt{ 4r^{2} - 8q^{2} + \sigma^{2}_{_{0}}  } }{ \sqrt{ 8q^{2} - \sigma^{2}_{_{0}}} } \right)
\label{phi_Maxwell} \\
\dot{\boldsymbol{f}}(\phi) &=&  
\frac{q^{2} (\sigma^{2}_{_{0}} - 8q^{2}) ~ \phi }{ 4\sigma^{2}_{_{0}} }\sec^{4}\!\left(\! \frac{ \sqrt{8q^{2}-\sigma^{2}_{_{0}}} ~\phi }{ 2\sigma_{_{0}} } \!\right).
\label{f_Maxwell} 
\end{eqnarray}   
Using these expression, the sGB coupling function $\boldsymbol{f}(\phi)=\int^{\phi}_{\phi_{_{0}}}\dot{\boldsymbol{f}}(\tilde{\phi})d\tilde{\phi}$ becomes
\begin{equation}\label{fphi_ellis}
\boldsymbol{f}\!(\phi) 
\!=\!\frac{q^{2}}{6}\!\!\left\{ \!\sec^{2}\!\!\left(\!\!\frac{\sqrt{8q^{2}\!-\!\sigma^{2}_{_{0}}}\!\!~\phi}{2\sigma_{_{0}}}\!\!\right)\!\!+\!\ln\!\!\left[\! \sec^{4}\!\!\left(\!\!\frac{\sqrt{8q^{2} \!-\! \sigma^{2}_{_{0}}}\!\!~\phi}{2\sigma_{_{0}}}\!\!\right)\!\right]\!-\!\frac{\sqrt{8q^{2} \!-\! \sigma^{2}_{_{0}}}\!\!\!~~\phi}{\sigma_{_{0}}} \!\!
\left[\!2\!+\!\sec^{2}\!\!\left(\!\!\frac{\sqrt{8q^{2} - \sigma^{2}_{_{0}}}\!\!~\phi}{2\sigma_{_{0}}}\!\!\right)\!\right]\!\!\tan\!\!\left(\!\!\frac{\sqrt{8q^{2} - \sigma^{2}_{_{0}}}\!\!~\phi}{2\sigma_{_{0}}}\!\!\right)  
\!\right\} + \alpha_{_{(2)}},
\end{equation}
where $\alpha_{_{(2)}}$ is a constant determined by the parameters $q$, $\sigma_{_{0}}$ and $\phi_{_{0}}=\phi(r_{_{0}})$. 
Without any loss of generality, one can always set
 $\alpha_{_{(2)}}=0$. It is important  to note that this first example corresponds to the magnetic dual of the  electrically charged Ellis-Bronnikov wormhole in EsGB Maxwell gravity, presented in Ref. \cite{Canate2019}. 

\begin{itemize}
\item[{\bf1.}]{\bf Magnetically charged Ellis-Bronnikov T-WH in EsGB Maxwell theory without exotic matter}
\end{itemize}
It can be seen that the line element (\ref{EsGB_maxwell}), for the case $8q^{2}>\sigma^{2}_{_{0}}$, has an  ultrastatic spherically symmetric and asymptotically flat WH structure with shape function $b_{_{_{\mathrm{LED}}}}\!\!(r) = (8q^{2}-\sigma^{2}_{_{0}}) /(4r)$, and throat radius given by 
$r_{_{0}} = \sqrt{(8q^{2}-\sigma^{2}_{_{0}})/4}$.
Moreover, since this shape function $b_{_{_{\mathrm{LED}}}}\!\!(r)$ satisfies the flaring-out condition, that is 
\begin{equation}
1 - \frac{b_{_{_{\mathrm{LED}}}}\!\!(r)}{r} > 0 \quad\quad\quad \forall~ r > r_{_{0}}, \quad\quad\quad\quad\textup{and}\quad\quad\quad\quad b'_{_{_{\mathrm{LED}}}}\!\!(r_{_{0}}) = -1, 
\end{equation}
we conclude that, for this case, the line element admits a T-WH interpretation.
In contrast with the Ellis and Bronnikov works \cite{ellis73,Bronnikov73}, here the scalar field \eqref{phi_Maxwell} is real in the whole T-WH spacetime.
Furthermore, notice that the electromagnetic Lagrangian in  consideration, {\it i.e.} $\mathcal{L}_{_{_{\mathrm{LED}}}} = \mathcal{F} = \frac{q^{2}}{r^{4}}$, satisfies the constraints \eqref{NEC_NLED} and \eqref{WEC_NLED}, guaranteeing the positivity of the local energy density associated with the magnetic field. 

Summarizing, the metric (\ref{EsGB_maxwell}) with  $8q^{2}>\sigma^{2}_{_{0}}$ becomes a purely magnetic ultrastatic traversable wormhole solution supported by nonexotic matter in the framework of EsGB-Maxwell gravity.  
%
%
\begin{figure}
\centering
\epsfig{file=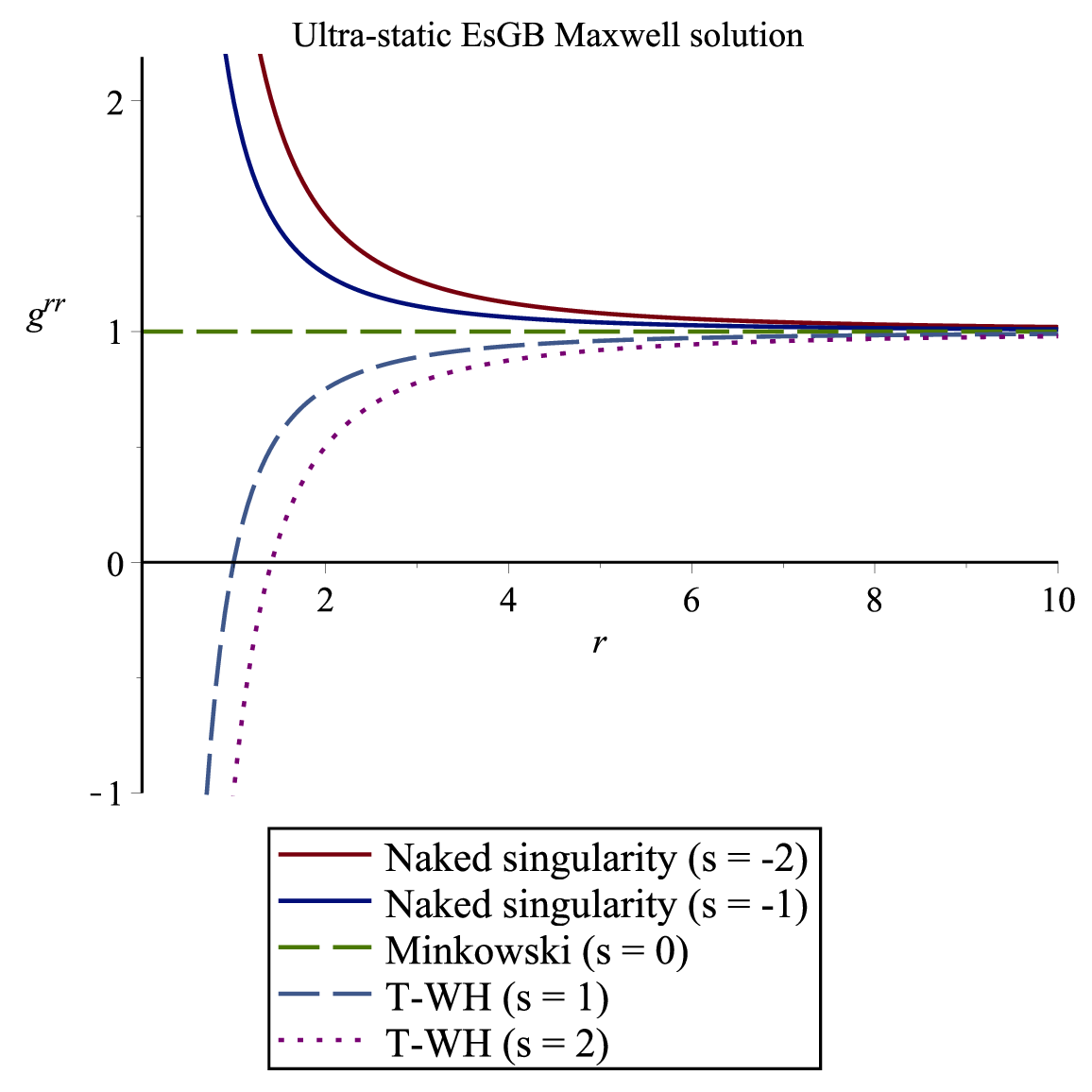, scale=0.44}
\caption{ \label{shapeMaxwell} Behavior of $g^{rr}= 1 - b_{_{_{\mathrm{LED}}}}\!\!(r)/r$, for different values of $s=(8q^{2}-\sigma^{2}_{_{0}})/4$. The ordinate is $g^{rr} = 1 - s/r^{2}$; T-WHs are only possible for $s>0$. The abscissa is $r$; zeros indicate the location of a wormhole throat.   
}
\end{figure}

Figure \ref{shapeMaxwell}  shows the behavior of the shape function for several values of $s=(8q^{2}-\sigma^{2}_{_{0}})/4$; for $s<0$ the solution describes a naked singularity spacetime; $s=0$, the solution becomes the Minkowski metric; $s>0$, the solution admits an  ultrastatic T-WH interpretation.    

Below, the particular cases of vanishing scalar field ($\sigma_{_{0}} = 0 \neq q$), and 
vanishing magnetic field ($\sigma_{_{0}} \neq 0 = q$) are discussed. 

\begin{itemize}
\item[{\bf2.}] {\bf Magnetically charged Ellis-Bronnikov T-WH in EGB Maxwell theory with variable GB coefficient}
\end{itemize}
The line element (\ref{EsGB_maxwell}) for the particular case with $\sigma_{_{0}} = 0 \neq q$ takes the form 
\begin{equation}\label{Ellis_Br}
ds^{2} = - dt^{2} + \left( 1 - \frac{2q^{2}}{r^{2}}  \right)^{\!\!^{-1}} 
dr^{2} + r^{2}(d\theta^{2}  + \sin^{2}\theta d\varphi^{2}),    
\end{equation}
%
which has a magnetically charged ultrastatic T-WH interpretation, since this line element is equivalent to an Ellis-Bronnikov wormhole, Eq. \eqref{EllisWH}, but with a throat of radius determined by the magnetic charge only, $r_{_{0}} = \sqrt{2 q^{2}}$. On the other hand, regarding the absence of scalar charge  
$\sigma_{_{0}} = 0$, this  implies that $\phi = 0$, $\phi' = 0$, $\phi'' = 0$.  However, according to Eqs. \eqref{phi_Maxwell} and \eqref{fphi_ellis}, the following limits $\lim\limits_{\sigma_{_{0}}\to0}\frac{~\phi~}{\sigma_{_{0}}}$  
and 
$\lim\limits_{\sigma_{_{0}}\to0}\boldsymbol{f}(\phi)$  
exist, and correspond to well-defined (nontrivial) functions of $r$. On the other hand, 
$\lim\limits_{\sigma_{_{0}}\to0}\dot{\boldsymbol{f}}(\phi) \rightarrow \infty$, and $\lim\limits_{\sigma_{_{0}}\to0}\ddot{\boldsymbol{f}}(\phi) \rightarrow \infty$.  
Despite this, the quantities $\dot{\boldsymbol{f}}(\phi)$ and $\ddot{\boldsymbol{f}}(\phi)$ appear in the field equations (\ref{Eqt}), (\ref{Eqr}) and (\ref{Eqte}), as multiplicative factors to derivatives of the scalar field, {\it i.e.}  $\phi'\dot{\boldsymbol{f}}(\phi)$, $\phi''\dot{\boldsymbol{f}}(\phi)$ and $\phi'^{2}\ddot{\boldsymbol{f}}(\phi)$, and these factors (as $\sigma_{_{0}}=0$) are well-defined functions in whole the wormhole domain, given by 
\begin{equation}\label{EspCa}
 \phi'\dot{\boldsymbol{f}}(\phi)\Big|_{\sigma_{_{0}}=0} = - \frac{  r^{3} \tan^{-1}\!\!\left(\sqrt{\frac{ r^{2} - 2q^{2}  }{2q^{2}}}\right)  }{ \sqrt{8}q\sqrt{r^{2} - 2q^{2} } }, \quad \phi''\dot{\boldsymbol{f}}(\phi) \Big|_{\sigma_{_{0}}=0} =  \frac{ \left( r^{2} - q^{2} \right)r^{2} \tan^{-1}\!\!\left(\sqrt{\frac{ r^{2} - 2q^{2}  }{2q^{2}}}\right)  }{ \sqrt{2} q  \left( r^{2} - 2q^{2} \right)^{\frac{3}{2}}  }
\end{equation}
and
\begin{equation}\label{EspCb}
 \phi'^{2}\ddot{\boldsymbol{f}}(\phi) \Big|_{\sigma_{_{0}}=0} = \frac{ r^{2} \left[ q + \sqrt{ 8(r^{2} - 2q^{2})}\tan^{-1}\!\!\left(\sqrt{\frac{ r^{2} - 2q^{2}  }{2q^{2}}}\right) \right] }{ 2 q \left( 2q^{2} - r^{2} \right)  }.
\end{equation}
Therefore there is no conflict with the fulfillment of the field equations when $\sigma_{_{0}}=0$.\\
Alternatively, the above can also be interpreted as follows: when $\sigma_{_{0}}=0$, the scalar field turns off $\phi(r) = 0$, and the EsGB theory defined by Eqs. \eqref{phi_Maxwell} and \eqref{fphi_ellis} reduces to an Einstein-Gauss-Bonnet theory with variable Gauss-Bonnet coefficient $\alpha_{_{GB}}(r) =\boldsymbol{f}(r)
=\lim\limits_{\sigma_{_{0}}\to0}\boldsymbol{f}(\phi)$ given by 
\begin{equation}\label{cfunct}
\boldsymbol{f}(r)
= \frac{r^{2}}{12} + \frac{q^{2}}{3}\ln\left(\frac{ r^{2} }{ q^{2} }\right) - \frac{1}{\sqrt{72}}\!\left(4q + \frac{r^{2}}{q}\right)\!\sqrt{ r^{2} - 2q^{2}   }\tan^{-1}\!\!\left(\sqrt{\frac{ r^{2} - 2q^{2}  }{2q^{2}}}\right),
\end{equation}
so that $\alpha'_{_{GB}}(r)= \phi'\dot{\boldsymbol{f}}(\phi)\Big|_{\sigma_{_{0}}=0}$ and $\alpha''_{_{GB}}(r)=\phi''\dot{\boldsymbol{f}}(\phi) \Big|_{\sigma_{_{0}}=0} + \phi'^{2}\ddot{\boldsymbol{f}}(\phi) \Big|_{\sigma_{_{0}}=0}$.
In summary, we can say that when $\sigma_{_{0}}=0$  the line element \eqref{EsGB_maxwell} is a purely magnetic solution of an EGB-Maxwell theory with variable Gauss-Bonnet coefficient, $\alpha_{_{GB}}=\boldsymbol{f}(r)$, given by \eqref{cfunct}.  In this case the traversable Ellis-Bronnikov wormhole is nonexotic, and is supported by the source-free magnetic field\footnote{That is, a magnetic field solution of the source-free generalized Maxwell equations \eqref{NLED_Eqs}. } and $\boldsymbol{f}$GB curvature. 

\begin{itemize}
\item[{\bf3.}]{\bf Ellis-Bronnikov wormhole without electric charge supported by a phantom scalar field}
\end{itemize}
For the particular case $\sigma_{_{0}} \neq 0 = q$, the line element (\ref{EsGB_maxwell}) becomes 
\begin{equation}\label{NoTWH}
ds^{2} = - dt^{2} + \left( 1 + \frac{\sigma^{2}_{_{0}}}{4r^{2}}  \right)^{\!\!^{-1}}dr^{2} + r^{2}(d\theta^{2}  + \sin^{2}\theta d\varphi^{2})    
\end{equation}
while the expressions (\ref{phi_Maxwell}) and (\ref{f_Maxwell}) take the form
\begin{equation}\label{NoTWH_sf}
\phi = \frac{2\sigma_{_{0}}}{\sqrt{ - \sigma^{2}_{_{0}} }}\tan^{-1}\left(  \frac{ \sqrt{ 4r^{2} + \sigma^{2}_{_{0}}  } }{ \sqrt{ - \sigma^{2}_{_{0}}} } \right) = \pm 2 \tanh^{-1}\left(   \sqrt{ \frac{ 4r^{2} + \sigma^{2}_{_{0}}  }{  \sigma^{2}_{_{0}}}  } \right), \quad\quad  
\dot{\boldsymbol{f}}(\phi) = 0. 
\end{equation}
Hence, we conclude that when $q=0$, the electromagnetic effects are turned off and the original Ellis-Bronnikov work \eqref{Lfcero_metric}$-$\eqref{SfactionL2} is recovered. 

\subsection{ New magnetically charged wormhole in EsGB Euler-Heisenberg gravity}
%
In the following we shall consider the Euler-Heisenberg nonlinear electrodynamics model in the approximation of the weak-field limit: 
\begin{equation}\label{EH}
\mathcal{L}_{_{_{\mathrm{EH}}}} 
= 
\mathcal{L}_{_{_{\mathrm{LED}}}} + \frac{ \gamma }{2}\mathcal{F}^{2}.  
\end{equation}
In order to expand the range of this model, let us consider $\gamma$ as an arbitrary parameter  which can be positive, negative or null. 
Notice that, regardless of the sign of the parameter $\gamma$,
the electrodynamics model \eqref{EH} satisfies the correspondence to Maxwell theory ({\it 
i.e.} $\mathcal{L}_{_{_{\mathrm{EH}}}} \approx \mathcal{F}$ as $\mathcal{F}\approx0$).
On the other hand, evaluating  
\eqref{Fsol} from \eqref{EH} and then substituting into \eqref{UltraS_Solution} we find  
the following novel ultrastatic metric:   
\begin{equation}\label{EsGB_EH}
ds^{2} =  - dt^{2} + \left( 1 - \frac{ 8q^{2}-\sigma^{2}_{_{0}} }{4r^{2}} - \frac{ \gamma q^{4}  
}{r^{6}} \right)^{\!\!^{-1}} dr^{2} + r^{2}(d\theta^{2}  + \sin^{2}\theta d\varphi^{2})
\end{equation}
while, according to (\ref{phi_sol}) and (\ref{df_sol}),  the corresponding EsGB model for which this metric is an exact purely magnetic solution of the EsGB Euler-Heisenberg  field equations can be determined by
\begin{eqnarray}
\phi(r)\!\!&=&\!\!\int^{r}_{r_{_{0}}} \frac{2 \sigma_{_{0}} \chi  }{ \sqrt{ 4\chi^{6} -(8q^{2}-\sigma^{2}_{_{0}})\chi^{4}   - 4 \gamma q^{4} }  } d\chi, \quad\quad\quad \boldsymbol{f}(r)\!\!=\!\!\int^{r}_{r_{_{0}}}\phi'(\tilde{r})\dot{\boldsymbol{f}}(\tilde{r})d\tilde{r}
\label{phi_EH} \\
\dot{\boldsymbol{f}}(r)\!\!&=&\!\!-\frac{3r^{6}}{2\sigma_{_{0}}\!\!\left[(8q^{2}-\sigma^{2}_{_{0}})r^{4} + 4 \gamma q^{4} \right]}\!
\!\left(\!\!  \sqrt{ 4r^{6} -(8q^{2}-\sigma^{2}_{_{0}})r^{4} - 4\gamma q^{4} } + \frac{2r^{2}}{3\sigma_{_{0}}}\int^{r}_{r_{_{0}}}(4q^{2} - 3\chi^{2})\phi'(\chi) d\chi \right). 
\label{f_EH} 
\end{eqnarray}

In accordance with Sec. \ref{Metodo}, one can see that the above expressions 
(\ref{EH})$-$(\ref{f_EH}) verify the EsGB-$\mathcal{L}(\mathcal{F})$ field equations (\ref{Eqt})$-$(\ref{phi2}) for the pure magnetic field.\footnote{That is, with 
electromagnetic field tensor of the form \eqref{magnetica}.}

{\it Limit case of zero scalar field charge.}  For the case $\sigma_{_{0}}=0$,  the line element \eqref{EsGB_EH} becomes
\begin{equation}\label{EsGB_EH_gr}
ds^{2} = - dt^{2} + \left( 1 - \frac{2q^{2}}{r^{2}} - \frac{ \gamma q^{4}  
}{r^{6}}  \right)^{\!\!^{-1}} dr^{2} + r^{2}(d\theta^{2}  + \sin^{2}\theta d\varphi^{2}).    
\end{equation}
For this case, according to Eq. \eqref{f_EH}, the scalar field becomes vanished 
$\phi = 0$ for all values of the $r$ coordinate. 
It is to be noted that in this case
the following expressions  
$\phi/\sigma_{_{0}}$, 
$\phi'\dot{\boldsymbol{f}}$, 
$\phi''\dot{\boldsymbol{f}}$ and 
$\phi'^{2}\ddot{\boldsymbol{f}}$, when  $\sigma_{_{0}}\rightarrow0$, become nontrivial well-defined functions
of the radial coordinate. For instance, as $\sigma_{_{0}}\to0$ the factor  $\phi'\!\dot{\boldsymbol{f}}$  becomes  
\begin{eqnarray}\label{phf_r_b}
\phi'(r)\!\dot{\boldsymbol{f}}(r)  
 \!\!&=&\!\! -\frac{3r^{7}}{\sqrt{ \!4r^{6} \!- \! 8q^{2}r^{4} \!-\!4 \gamma 
 q^{4} } \left[8q^{2}r^{4}\!+\!4\gamma q^{4} \right] } 
\!\bigg(\!\!\sqrt{ 4r^{6}\!-\!8q^{2}r^{4}\!-\!4\gamma q^{4} }  
+\!\frac{4r^{2}}{3}\!\!\int^{r}_{r_{_{0}}}\!\!\!   
\frac{(4q^{2}\!-\!3\chi^{2})\chi}{ \sqrt{ 4\chi^{6}\!-\!8q^{2}\chi^{4}\!-\!4 \gamma q^{4} }  }  
d\chi \!\bigg)
\end{eqnarray}
and then $(\phi'\!\dot{\boldsymbol{f}})' = \phi''\!\dot{\boldsymbol{f}} + (\phi')^{2}\!\ddot{\boldsymbol{f}}$ is well defined for the case with $\sigma_{_{0}} = 0$.
Summarizing, we can say that, when $\sigma_{_{0}}=0$, the scalar field becomes trivial, $\phi(r)=0$. However, the line element (\ref{EsGB_EH_gr}) is still a pure-magnetic ultrastatic solution of the EsGB Euler-Heisenberg system. Alternatively, this frame of EsGB-Heisenberg gravity is equivalent to an EGB Euler-Heisenberg theory 
with a variable GB coefficient determined by $\alpha_{_{GB}}(r) = \int^{r}_{r_{_{0}}}\phi'(\tilde{r})\dot{\boldsymbol{f}}(\tilde{r})d\tilde{r}$ with $\phi'(r)\dot{\boldsymbol{f}}(r)$
given by \eqref{phf_r_b}. 


{\it Limit case of linear electrodynamics.} For the case $\gamma = 0$ (for all $\sigma_{_{0}}$, $q$) the Lagrangian density \eqref{EH} becomes the Maxwell's electrodynamics Lagrangian, the metric \eqref{EsGB_EH} reduces to \eqref{EsGB_maxwell}, and the scalar field \eqref{phi_EH} becomes \eqref{phi_Maxwell}; whereas the expression \eqref{f_EH} reduces to \eqref{f_Maxwell}. Therefore, for this limit case the solution \eqref{EsGB_maxwell}$-$\eqref{fphi_ellis} is recovered. \\
Now we consider the particular cases $\gamma>0$ and $\gamma<0$,
and we determine the parameter settings needed so that the three-parameter solution  
\eqref{EsGB_EH} admits T-WH interpretations. \\

\begin{figure}
\centering
\epsfig{file=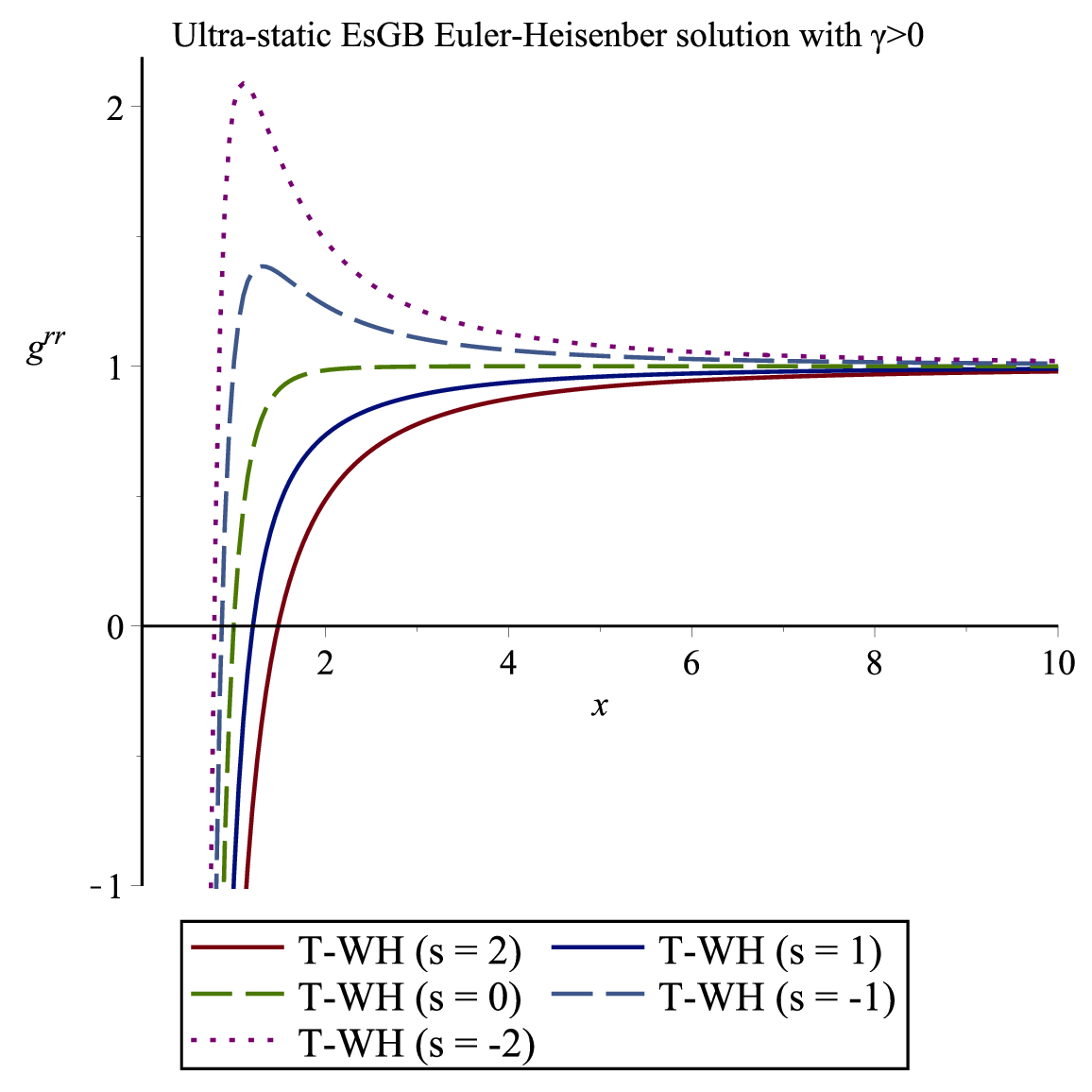, scale=0.44}
\caption{ \label{shapeEH} Behavior of $g^{rr} = 1 - b_{_{_{\mathrm{EH}}}}\!\!(r)/r$, for several real  
values of the parameter $s=a_{_{0}}/a^{^{\frac{1}{3}}}_{_{1}}$, with $a_{_{0}}=(8q^{2}-\sigma^{2}_{_{0}})/4 \in \mathbb{R}$ and $a_{_{1}}=\mu^{2}q^{4}\in \mathbb{R}^{+}\!-\!\{0\}$, is shown; the ordinate is $g^{rr} = 1 - s/x^{2} - 1/x^{6}$; the abscissa is $x$ being $x=r/a^{^{\frac{1}{6}}}_{_{1}}$.}
\end{figure}

\begin{center}
{\it 1. Case $\boldsymbol{\gamma>0}$}
\end{center}
For this case replacing $\gamma = \mu^{2}$ (being $\mu\in\mathbb{R}$) into Eq.\eqref{EH}, and since $\mathcal{F}=q^{2}/r^{4}$ ({\it i.e.} positive defined for any values of the radial coordinate), yields that $\mathcal{L}_{_{_{\mathrm{EH}}}}$ obeys the following inequalities: 
\begin{equation}\label{EH_wecA}
 \mathcal{L}_{_{_{\mathrm{EH}}}} = \mathcal{F} + \frac{\mu^{2}}{2}\mathcal{F}^{2} \geq 0 \quad\quad\textup{ and } \quad\quad \frac{d\mathcal{L}_{_{_{\mathrm{EH}}}}}{d\mathcal{F}} = 1 + \mu^{2}\mathcal{F} > 0.  
\end{equation}
This indicates that the energy momentum tensor associated with this nonlinear electrodynamics satisfies the null and weak energy conditions \eqref{NEC_NLED} and \eqref{WEC_NLED}. 
Furthermore, the line element \eqref{EsGB_EH} takes the form
\begin{equation}\label{EsGB_EHgMa}
ds^{2} =  - dt^{2} + \left( 1 - \frac{ 8q^{2}-\sigma^{2}_{_{0}} }{4r^{2}} - \frac{ \mu^{2} q^{4} }{r^{6}} \right)^{\!\!^{-1}} dr^{2} + r^{2}(d\theta^{2}  + \sin^{2}\theta d\varphi^{2})
\end{equation}
which has an ultrastatic WH structure with
shape function given by $b_{_{_{\mathrm{EH}}}}\!\!(r) = \frac{ 8q^{2}-\sigma^{2}_{_{0}} }{4r} + \frac{\mu^{2} q^{4}}{r^{5}}$.
For simplicity $b_{_{_{\mathrm{EH}}}}\!\!(r) = a_{_{0}}/r + a_{_{1}}/r^{5}$, with $a_{_{0}}=(8q^{2}-\sigma^{2}_{_{0}})/4$ and $a_{_{1}}=\mu^{2}q^{4}$. 
Specifically, considering nontrivial Euler-Heisenberg  effects ({\it i.e.} imposing that  $a_{_{1}}\neq0$) yields 
\begin{equation}
g^{rr} = 1 - \frac{b_{_{_{\mathrm{EH}}}}\!\!(r)}{r}  
= 1 - \frac{s}{x^{^{2}}} - \frac{1}{x^{^{6}}},   
\quad\quad\quad s =  \frac{a_{_{0}}}{a^{^{\frac{1}{3}}}_{_{1}}}, \quad\quad\quad x = \frac{r}{a^{^{\frac{1}{6}}}_{_{1}}}.
\end{equation}
where $x$ is an auxiliary variable with range $x\in(0,\infty)$.
In Fig. \ref{shapeEH} we can see that in the whole range  
of auxiliary variable $x$, the function $g^{rr}(x)$ vanishes only at
$x=x_{_{0}}$, where $x_{_{0}}$ is such that
\begin{equation}
x^{2}_{_{0}} = \frac{r^{2}_{_{0}}}{a^{^{\frac{1}{3}}}_{_{1}}} \!=\! \frac{1}{3}\!\!\left(\!\frac{a_{_{0}}}{a^{^{\frac{1}{3}}}_{_{1}}}\!\right) +  
\frac{1}{6}\!\!\left[\!108\!+\!\frac{8a^{3}_{_{0}}}{a_{_{1}}}\!+\!12\!\left(\frac{12a^{3}_{_{0}}}{a_{_{1}}}\!+\! 81\!\right)^{\!\!\frac{1}{2}}\!\right]^{\!\frac{1}{3}}\!\!\!+ \frac{2}{3}\!\!\left(\!\frac{a_{_{0}}}{a^{^{\frac{1}{3}}}_{_{1}}}\!\right)^{\!\!\!2}\!\left[\!108 \!+\!\frac{8a^{3}_{_{0}}}{a_{_{1}}}\!+\! 12\left[ \frac{12a^{3}_{_{0}}}{a_{_{1}}}\!+\!81\right]^{\!\frac{1}{2}} \right]^{\!-\frac{1}{3}}  
\end{equation}
Then, the throat radius squared is given by
\begin{equation}
r^{2}_{_{0}} = \frac{1}{3}a_{_{0}} + \frac{1}{6}\left[108a_{_{1}} + 8a^{3}_{_{0}} + 12\left(12a^{3}_{_{0}}a_{_{1}} + 81a^{2}_{_{1}} \right)^{\frac{1}{2}} \right]^{\frac{1}{3}} + \frac{2}{3}a^{2}_{_{0}}\left[108a_{_{1}} + 8a^{3}_{_{0}} + 12\left(12a^{3}_{_{0}}a_{_{1}} + 81a^{2}_{_{1}} \right)^{\frac{1}{2}} \right]^{-\frac{1}{3}}
\end{equation}
which satisfies 
\begin{equation}
1 - \frac{b_{_{_{\mathrm{EH}}}}\!\!(r)}{r} = 1 - \frac{a_{_{0}}}{r^{2}} - \frac{a_{_{1}}}{r^{6}} > 0 \quad\quad \forall~ r \geq r_{_{0}},  
\quad\quad b'_{_{_{\mathrm{EH}}}}\!\!(r_{_{0}}) = -\frac{a_{_{0}}}{r^{2}_{_{0}}} - \frac{5a_{_{1}}}{r^{6}_{_{0}}} = -\frac{a_{_{0}}}{r^{2}_{_{0}}} - \frac{a_{_{1}}}{r^{6}_{_{0}}} - \frac{4a_{_{1}}}{r^{6}_{_{0}}} = -1- \frac{4a_{_{1}}}{r^{6}_{_{0}}} < 1.
\end{equation}
Thus the solution \eqref{EsGB_EH}, under the action of nontrivial  
effects of EH theory ({\it i.e.} $\mathcal{L}_{_{\mathrm{EH}}}$ with $\mu$, $q$ $\in \mathbb{R} - \{0\}$), fulfills the flaring out condition, and hence represents a traversable wormhole interpretation which is supported by nonexotic matter, since EH theory satisfies the WEC Eq. \eqref{EH_wecA} and the scalar field \eqref{phi_EH} is real in the whole T-WH spacetime, that is  
\begin{equation}
\phi(r) = \int^{r}_{r_{_{0}}} \frac{ \sigma_{_{0}}  }{ \chi^{2} \sqrt{ 1 - \frac{a_{_{0}}}{ \chi^{2}}   -  \frac{a_{_{1}}}{ \chi^{6}} }  } d\chi \geq 0 \quad\quad\quad\quad \forall~ r \geq r_{_{0}}.
\end{equation}
Hence, for the setting of parameters $\{ \gamma>0, \quad  q\in \mathbb{R}-\{0\}, \quad \sigma_{_{0}} \in \mathbb{R}  \}$ the line element \eqref{EsGB_EH_gr} can be interpreted as the metric of a T-WH spacetime supported by nonexotic matter in the gravity context of EsGB Euler-Heisenberg gravity.

{\bf New magnetically charged T-WH in EGB Euler-Heisenberg ($\boldsymbol{\gamma>0}$) theory with variable $\boldsymbol{\alpha_{{GB}}}$. } 
The line element \eqref{EsGB_EHgMa} for the particular case $\sigma_{_{0}} = 0$ reduces to
\begin{equation}\label{EsGB_EH_b}
ds^{2} = - dt^{2} + \left( 1 - \frac{2q^{2}}{r^{2}} - \frac{\mu^{2} q^{4}  
}{r^{6}}  \right)^{\!\!^{-1}}dr^{2} + r^{2}(d\theta^{2}  + \sin^{2}\theta d\varphi^{2})    
\end{equation}
which still admits a T-WH interpretation. We can say that when $\sigma_{_{0}}=0$, the scalar field becomes trivial, $\phi(r)=0$. However, the line element (\ref{EsGB_EH_b}) 
is a pure-magnetic T-WH solution of an EGB Euler-Heisenberg theory with variable GB coefficient 
$\alpha_{_{GB}}(r)=\boldsymbol{f}(r)=\int^{r}_{r_{_{0}}}\phi'(\tilde{r})\dot{\boldsymbol{f}}(\tilde{r})d\tilde{r}$, where $\phi'(r)\dot{\boldsymbol{f}}(r)$ is determined by (\ref{phf_r_b}) with $\gamma=\mu^{2}$.
In this case the traversable wormhole is supported by a physically reasonable Euler-Heisenberg
electrodynamics and $\boldsymbol{f}$GB curvature. \\  

\begin{figure}
\centering
\epsfig{file=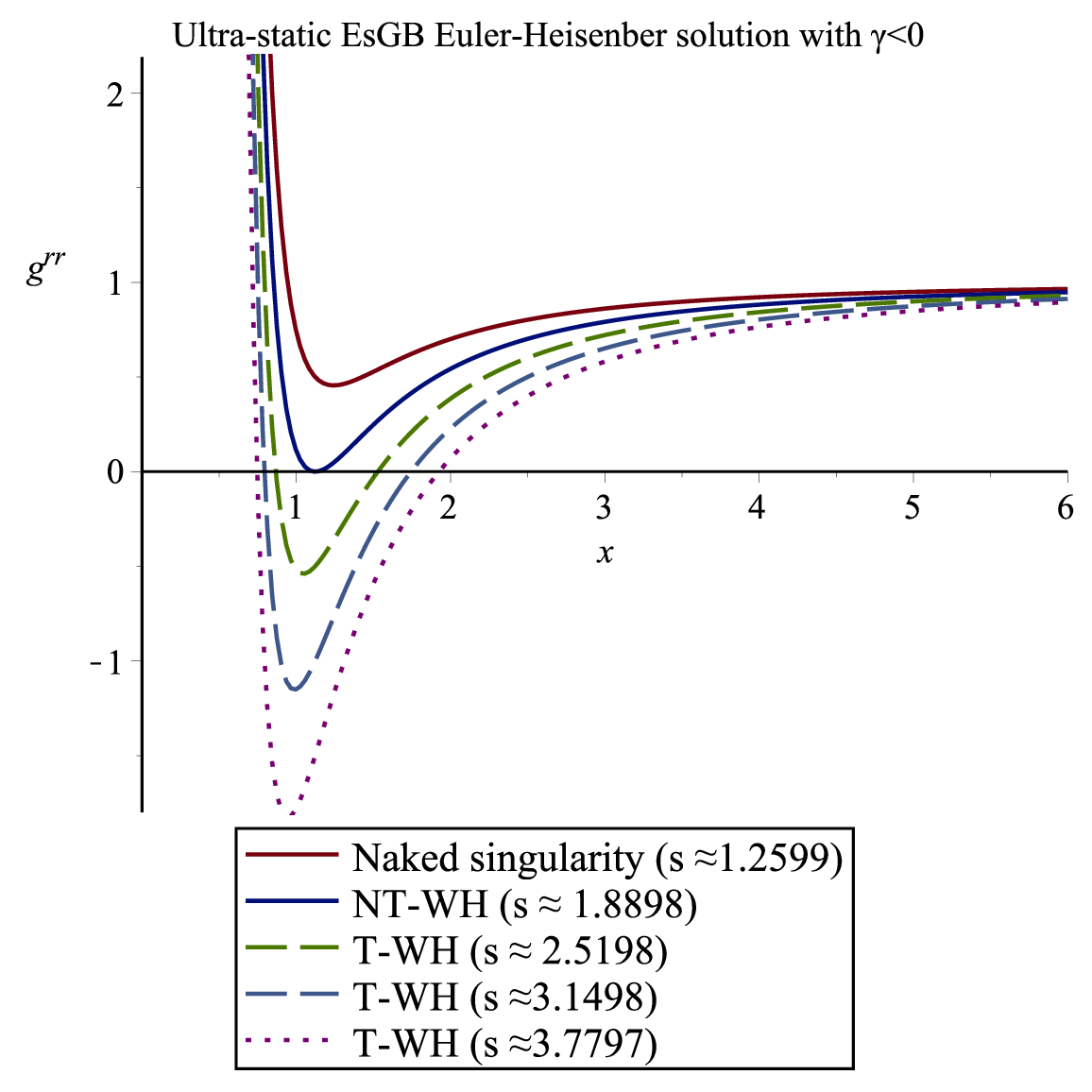, scale=0.44}
\caption{ \label{shapeEH2} Behavior of $g^{rr} = 1 - b_{_{_{\mathrm{EH}}}}\!\!(r)/r$, for several real 
values of the parameter $s=a_{_{0}}/a^{^{\frac{1}{3}}}_{_{1}}$ [with $a_{_{0}}=(8q^{2}-\sigma^{2}_{_{0}})/4 \in \mathbb{R}$ and $a_{_{1}}=\mu^{2}q^{4}\in \mathbb{R}^{+}\!-\!\{0\}$], is shown; the ordinate is $g^{rr} = 1 - s/x^{2} + 1/x^{6}$; the abscissa is $x$ being $x=r/a^{^{\frac{1}{6}}}_{_{1}}$.  
}
\end{figure}

\begin{center}
{\it 2. Case $\boldsymbol{\gamma<0}$}
\end{center}
For this case replacing $\gamma = -\mu^{2}$ (being $\mu\in\mathbb{R}$) into Eq.\eqref{EH} yields
\begin{equation}\label{EH_wec}
 \mathcal{L}_{_{_{\mathrm{EH}}}} = \mathcal{F} - \frac{\mu^{2}}{2}\mathcal{F}^{2} = \frac{ q^{2} }{ r^{4} } \left(1 - \frac{ \mu^{2} q^{2} }{ 2 r^{4} } \right) \quad\textup{ and } \quad \frac{d\mathcal{L}_{_{_{\mathrm{EH}}}}}{d\mathcal{F}} = 1 - \mu^{2}\mathcal{F} = 1 - \frac{ \mu^{2} q^{2} }{  r^{4} }. 
\end{equation}
Hence, we conclude that the energy momentum tensor associated with this nonlinear electrodynamics
satisfies the null and weak energy condition ({\it i.e.}  
the inequalities $\mathcal{L}_{_{_{\mathrm{EH}}}}\geq0$ and $\frac{d\mathcal{L}_{_{_{\mathrm{EH}}}}}{d\mathcal{F}}\geq0$, are simultaneously satisfied) only for  
\begin{equation}\label{EH_n_wec}
r \geq \sqrt{|\mu q|}.    
\end{equation}
The line element \eqref{EsGB_EH} takes the form
\begin{equation}\label{EsGB_EHgMe}
ds^{2} =  - dt^{2} + \left( 1 - \frac{ 8q^{2}-\sigma^{2}_{_{0}} }{4r^{2}} + \frac{\mu^{2} q^{4}}{r^{6}} \right)^{\!\!^{-1}} dr^{2} + r^{2}(d\theta^{2}  + \sin^{2}\theta d\varphi^{2})
\end{equation}
with shape function given  by $b_{_{_{\mathrm{EH}}}}\!\!(r) = \frac{ 8q^{2}-\sigma^{2}_{_{0}} }{4r} - \frac{\mu^{2} q^{4}}{r^{5}}$.
For simplicity $b_{_{_{\mathrm{EH}}}}\!\!(r) = a_{_{0}}/r - a_{_{1}}/r^{5}$, with $a_{_{0}}=(8q^{2}-\sigma^{2}_{_{0}})/4$ and $a_{_{1}}=\mu^{2}q^{4}$.  
Specifically, considering nontrivial NLED  effects ({\it i.e.} imposing that $a_{_{1}}\neq0$) yields 
\begin{equation}
g^{rr} = 1 - \frac{b_{_{_{\mathrm{EH}}}}\!\!(r)}{r}  
= 1 - \frac{s}{x^{^{2}}} + \frac{1}{x^{^{6}}},   
\quad\quad\quad s =  \frac{a_{_{0}}}{a^{^{\frac{1}{3}}}_{_{1}}}, \quad\quad\quad x = \frac{r}{a^{^{\frac{1}{6}}}_{_{1}}}.
\end{equation}
We can see that the equation $1 - b_{_{_{\mathrm{EH}}}}\!\!(x_{_{0}})/x_{_{0}} = 0$ only admits real solutions
if $s \geq 3/(2)^{2/3}$, which are at most two positive solutions, $x_{_{out}}$ and $x_{_{int}}$, such that $x_{_{out}} \geq x_{_{int}}$, with $x^{2}_{_{out}}$ given by
\begin{equation}
x^{2}_{_{out}} = \frac{s}{3} + \frac{1}{6}\left[-108 + 8s^{3} + 12\left(-12s^{3} + 81\right)^{\!\frac{1}{2}}\right]^{\!\frac{1}{3}} + \frac{2}{3}s^{2}\!\left[ -108 + 8s^{3} + 12\left(-12s^{3} + 81\right)^{\!\frac{1}{2}}\right]^{\!-\frac{1}{3}}.   
\end{equation}
If $s=3/(2)^{2/3}$ this yields $x_{_{out}}=x_{_{int}}=2^{1/3}$, whereas, if $s > 3/(2)^{2/3}$ these solutions are such that $x_{_{out}} > 2^{1/3} > x_{_{int}}$, see  Fig. \ref{shapeEH2} for illustration.
%
%
Hence, for the case $s=3/(2)^{2/3}$ the solution \eqref{EsGB_EH} describes a 
wormhole spacetime with wormhole throat radius $r_{_{0}} = a^{^{\frac{1}{6}}}_{_{1}}x_{_{0}}$, 
with $x_{_{0}}=x_{_{out}}=2^{1/3}$, and satisfying $b'_{_{_{\mathrm{EH}}}}\!\!(r_{_{0}}) =1$ (indicating that  
the flaring out condition is not fulfilled) and hence for this particular case the corresponding  
metric describes a nontraversable wormhole spacetime (NT-WH). 
For the case  
$s > 3/(2)^{2/3}  
\approx 1.8898$, the wormhole solution has  
a wormhole throat radius $r_{_{0}} = a^{^{\frac{1}{6}}}_{_{1}}x_{_{0}}$, with $x_{_{0}} = x_{_{out}}> 2^{1/3}$,  
and fulfills the flaring out condition: 
\begin{equation}
1 - \frac{b_{_{_{\mathrm{EH}}}}\!\!(r)}{r} = 1 - \frac{a_{_{0}}}{r^{2}} + \frac{a_{_{1}}}{r^{6}} > 0 \quad\quad \forall~ r > r_{_{0}}, 
\quad\quad b'_{_{_{\mathrm{EH}}}}\!\!(r_{_{0}}) = -\frac{a_{_{0}}}{r^{2}_{_{0}}} + \frac{5a_{_{1}}}{r^{6}_{_{0}}} = -\frac{a_{_{0}}}{r^{2}_{_{0}}} + \frac{a_{_{1}}}{r^{6}_{_{0}}} + \frac{4a_{_{1}}}{r^{6}_{_{0}}} = -1 + \frac{4a_{_{1}}}{r^{6}_{_{0}}} < 1.
\end{equation}
Hence, for this case the solution \eqref{EsGB_EH} represents a traversable wormhole interpretation.
Since $x_{_{0}} = r_{_{0}}/a^{^{\frac{1}{6}}}_{_{1}} > 2^{1/3}$, this yields $r^{6}_{_{0}} > 4a_{_{1}}$.  
Thus, for the T-WH case $s > 3/(2)^{2/3}$, the domain of the radial coordinate $r\geq r_{_{0}}$ is restricted by a wormhole throat radius such that
\begin{equation}\label{r_02}
    r^{2}_{_{0}} >  \left(\frac{ 4|q|}{|\mu|}\right)^{^{\frac{1}{3}}}  |\mu q|. 
\end{equation}
Therefore, according to Eqs. \eqref{EH_n_wec} and \eqref{r_02}, by imposing  
$\left(\frac{4|q|}{|\mu|} \geq 1\right)$ the traversability of the wormhole without the requirement of
exotic mater in this modified gravity context is guaranteed.\\
Thus, for the setting of parameters $\{ \gamma <0, \quad  q\in \mathbb{R}-\{0\}, \quad \sigma_{_{0}} \in \mathbb{R}, \quad s > \frac{3}{(2)^{^{\frac{2}{3}}}}, \quad  \frac{4|q|}{|\mu|} \geq 1  \}$ the line element \eqref{EsGB_EH} can be interpreted as the metric of a T-WH spacetime supported by nonexotic matter in the gravity context of EsGB Euler-Heisenberg. 

{\bf New magnetically charged T-WH in EGB Euler-Heisenberg ($\boldsymbol{\gamma<0}$) theory with variable $\boldsymbol{\alpha_{{GB}}}$. } 
For the case of trivial scalar field $\sigma_{_{0}}=0$, the metric \eqref{EsGB_EHgMe} reduces to  
\begin{equation}\label{EsGB_EH_c}
ds^{2} = - dt^{2} + \left( 1 - \frac{2q^{2}}{r^{2}} + \frac{\mu^{2} q^{4}  
}{r^{6}}  \right)^{\!\!^{-1}} 
dr^{2} + r^{2}(d\theta^{2}  + \sin^{2}\theta d\varphi^{2}),    
\end{equation}
which, as long as $s>3/(2)^{2/3}$ and $4|q|/|\mu|\geq 1$,  
becomes a magnetically charged ultrastatic T-WH solution of an EGB Euler-Heisenberg theory with variable GB coefficient $\alpha_{_{GB}}(r)=\boldsymbol{f}(r)=\int^{r}_{r_{_{0}}}\phi'(\tilde{r})\dot{\boldsymbol{f}}(\tilde{r})d\tilde{r}$, where $\phi'(r)\dot{\boldsymbol{f}}(r)$ is given by \eqref{phf_r_b} with $\gamma=-\mu^{2}$.
This traversable wormhole is supported by a physically reasonable Euler-Heisenberg electrodynamics and $\boldsymbol{f}$GB curvature.


\subsection*{ \bf 3. New magnetically charged wormhole in EsGB Born-Infeld gravity }

Let us consider the Born-Infeld nonlinear electrodynamics model:
\begin{equation}\label{BI}
\mathcal{L}_{_{\mathrm{BI}}}
= 4\beta^{2} \left( -1 + \sqrt{ 1 + \frac{\mathcal{F}}{2\beta^{2}} }~\right),
\end{equation}
where  $\beta$ is a constant which has the physical interpretation of a critical field
strength \cite{BI}. \\
For this case, evaluating Eqs. \eqref{Fsol} and \eqref{UltraS_Solution} yields the following
ultrastatic metric: 
\begin{equation}\label{EsGB_BI}
ds^{2} =  - dt^{2} + \left[ 1 + \frac{ \sigma^{2}_{_{0}} }{ 4r^{2} } + 8\beta^{2} \left( r^{2}- \sqrt{ r^{4} + \frac{q^{2}}{2\beta^{2}} } ~ \right) \right]^{\!\!^{-1}} dr^{2}
+ r^{2}(d\theta^{2}  + \sin^{2}\theta d\varphi^{2}).
\end{equation}
According to (\ref{phi_sol}) and (\ref{df_sol}), the corresponding EsGB model for which this metric is an exact purely magnetic solution of the EsGB Euler-Heisenberg  field equations can be determined by
\begin{eqnarray}
\phi(r)\!\!&=&\!\!\int^{r}_{r_{_{0}}}{ \frac{ 2 \sigma_{_{0}}  }{  \sqrt{ 4\chi^{4} +  \sigma^{2}_{_{0}}\chi^{2} + 32\beta^{2} \chi^{4} \left( \chi^{2}- \sqrt{ \chi^{4} + \frac{q^{2}}{2\beta^{2}} } ~ \right) }  } d\chi}, \quad\quad\quad \boldsymbol{f}(r)\!=\!\int^{r}_{r_{_{0}}}\phi'(\tilde{r})\dot{\boldsymbol{f}}(\tilde{r})d\tilde{r} 
\label{phiBI}
\\
\dot{\boldsymbol{f}}(r)\!\!&=&\!\!\frac{ 16 \beta^{2} r^{4} }{\sigma^{2}_{_{0}}\!\!\left[\sigma^{2}_{_{0}} + 32\beta^{2}\left( r^{2} - \sqrt{ r^{4} + \frac{q^{2}}{2\beta^{2}} } \right)r^{2} \right]}\! \!
\int^{r}_{r_{_{0}}}\chi^{4}\left( 1 - \frac{\chi^{2}}{ \sqrt{ \chi^{4} + \frac{q^{2}}{2\beta^{2}} } }  \right)\phi'(\chi) d\chi. \label{pfBI}
\end{eqnarray}
{\it Limit case of linear electrodynamics.} For any $\sigma_{_{0}}$ and $q$, when $\beta\rightarrow \infty$,  the Lagrangian density \eqref{BI} becomes the Maxwell's electrodynamics Lagrangian, the scalar field \eqref{phiBI} becomes \eqref{phi_Maxwell},
whereas Eq. \eqref{pfBI} will be reduced to Eq. \eqref{f_Maxwell}. Therefore, for this limit case the solution \eqref{EsGB_maxwell}$-$\eqref{fphi_ellis} is recovered.

\begin{figure}
\centering
\epsfig{file=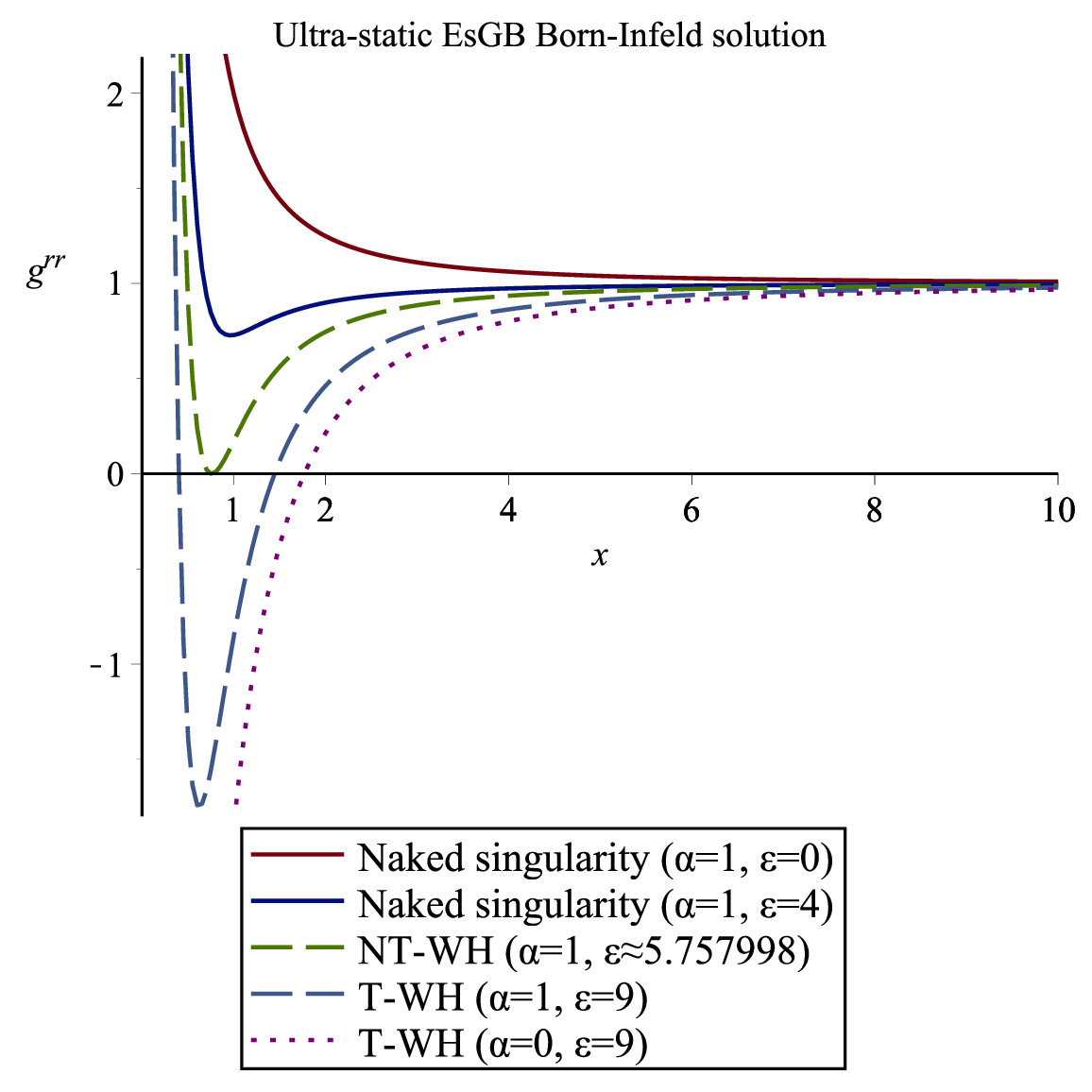, scale=0.44}
\caption{ \label{shapeBI} Behavior of $g^{rr}=1/g_{rr}$ with respect to $x$ 
 for different values of $\alpha =\frac{ \sigma^{2}_{_{0}} |\beta| }{ 4 |q|}$ and $\varepsilon =\frac{8|q\beta|}{\sqrt{2}}$. The ordinate is $g^{rr} = 1 + \alpha/x^{2} +  \varepsilon \left( \sqrt{2}x^{2} - \sqrt{ 2x^{4} + 1 ~} ~ \right)$; the abscissa is $x=\sqrt{\frac{|\beta|}{|q|} }~r$. 
}
\end{figure}

{\bf New magnetically charge T-WH in EsGB Born-Infeld theory without exotic matter.} 
For the solution \eqref{BI}$-$\eqref{pfBI}, the electromagnetic invariant $\mathcal{F}$ is positive definite $\mathcal{F}=\frac{q^{2}}{r^{4}} \geq 0$, therefore it is fulfilled
\begin{equation}
\mathcal{L}_{_{\mathrm{BI}}} 
= 4\beta^{2} \left( -1 + \sqrt{ 1 + \frac{\mathcal{F}}{2\beta^{2}}  
}~\right) \geq 0 \quad \textup{ and } \quad \frac{d\mathcal{L}_{_{\mathrm{BI}}}}{d\mathcal{F}} 
= \frac{1}{\sqrt{ 1 + \frac{\mathcal{F}}{2\beta^{2}} }} > 0    
\end{equation}
indicating that the the energy momentum tensor associated with this nonlinear electrodynamics satisfies the null and weak energy conditions \eqref{NEC_NLED}$-$\eqref{WEC_NLED}.
Moreover, the spacetime metric \eqref{EsGB_BI}
has an ultrastatic WH structure with  shape function given by  
$b_{_{_{\mathrm{BI}}}}\!\!(r) = -\frac{ \sigma^{2}_{_{0}} }{ 4r } - 8\beta^{2} \left( r^{3}- \sqrt{ r^{6} + \frac{q^{2}r^{2}}{2\beta^{2}} } ~ \right)$. Specifically, considering  
nontrivial Born-Infeld effects ({\it i.e.} as long as $q\neq0\neq\beta$), the metric component $g^{rr}= 1 - b_{_{_{\mathrm{BI}}}}\!\!(r)/r$ can be written as  
\begin{equation}
g^{rr} =  1 + \frac{\alpha}{x^{2}} +  \varepsilon \left( \sqrt{2}x^{2} - \sqrt{ 2x^{4} + 1 ~} ~ \right), 
\end{equation}
where $\alpha = \frac{ \sigma^{2}_{_{0}} |\beta| }{ 4 |q|}$, $\varepsilon =\frac{8|q\beta|}{\sqrt{2}}$ and $x=\sqrt{\frac{|\beta|}{|q|}}~r$.    
Plots of $g^{rr}$ as a function of 
 $x$, for different values of $\alpha$ and $\varepsilon$,
are illustrated in Fig. \ref{shapeBI}.  
\\
On the other hand, introducing the auxiliary parameter $\ell$ as
\begin{eqnarray}
\ell &=& 8 \alpha \varepsilon \!\left[ 3\!\left(1-\varepsilon^{2}\right)\!\!\left(\varepsilon^{2}+2\right) + \left(6 
\sqrt{2}\varepsilon^{2} - 8\alpha \varepsilon - 15\sqrt{2}
\right)\alpha \varepsilon \right] + 2\sqrt{2} \varepsilon^{2} (\varepsilon^{4} - 3\varepsilon^{2} + 3) - 2\sqrt{2} \nonumber\\ 
&& + 24\alpha \varepsilon^{2}\sqrt{ 
 6\sqrt{2}\left( 8\alpha^{2} + 3\epsilon^{2} - 6\sqrt{2}\alpha\varepsilon  + 5 \right)\alpha\varepsilon - 3(\varepsilon^{2}-1)^{^{2}} - 6\alpha^{2} } \!~,    
\end{eqnarray}
we conclude that only if the following expression is true, 
\begin{equation}\label{ell_imag}
\frac{ \ell^{^{\frac{1}{3}}} }{2\varepsilon} - \frac{ \varepsilon^{^{4}} - (32)^{^{\frac{1}{2}}} \alpha \varepsilon^{^{3}} + 8 \alpha^{^{2}} \varepsilon^{^{2}}  - (128)^{^{\frac{1}{2}}}\alpha \varepsilon - 2\varepsilon^{^{2}} + 1 }{\varepsilon\ell^{^{\frac{1}{3}}}} \in  \mathbb{I},  
\end{equation}
the equation $1 - b_{_{_{\mathrm{BI}}}}\!\!(x_{_{0}})/x_{_{0}} = 0$ admits positive real solutions. In fact, in general, there are at most two real positive solutions $x_{_{out}}$ and $x_{_{int}}$ such that $x_{_{out}}\geq x_{_{int}}$, respectively given by
\begin{eqnarray}
x^{2}_{_{out}} &=& \frac{ \ell^{^{\frac{1}{3}}} }{12\varepsilon} + \frac{ \varepsilon^{^{4}} - (32)^{^{\frac{1}{2}}} \alpha \varepsilon^{^{3}} + 8 \alpha^{^{2}} \varepsilon^{^{2}}  - (128)^{^{\frac{1}{2}}}\alpha \varepsilon - 2\varepsilon^{^{2}} + 1 }{6\varepsilon\ell^{^{\frac{1}{3}}}} +  \frac{ \varepsilon^{^{2}} - (8)^{^{\frac{1}{2}}} \alpha \varepsilon - 1}{ (72)^{^{\frac{1}{2}}} \varepsilon } \\
x^{2}_{_{int}} &=& -\frac{x^{2}_{_{out}}}{2} + \frac{ \varepsilon^{^{2}} - (8)^{^{\frac{1}{2}}} \alpha \varepsilon - 1}{ (32)^{^{\frac{1}{2}}} \varepsilon } - \frac{i}{(48)^{^{\frac{1}{2}}}} \left(  \frac{ \ell^{^{\frac{1}{3}}} }{2\varepsilon} - \frac{ \varepsilon^{^{4}} - (32)^{^{\frac{1}{2}}} \alpha \varepsilon^{^{3}} + 8 \alpha^{^{2}} \varepsilon^{^{2}}  - (128)^{^{\frac{1}{2}}}\alpha \varepsilon - 2\varepsilon^{^{2}} + 1 }{\varepsilon\ell^{^{\frac{1}{3}}}}  \right). 
\end{eqnarray}
Then, if $x^{2}_{_{out}} > x^{2}_{_{int}}$, the metric admits a T-WH interpretation with a wormhole throat radius $r_{_{0}} = x_{_{out}}$ holding $b'_{_{_{\mathrm{BI}}}}\!\!(r_{_{0}}) < 1$. If $x^{2}_{_{out}} = x^{2}_{_{int}} \in \mathbb{R}^{+}$ the metric admits a NT-WH interpretation with a wormhole throat radius $r_{_{0}} = x_{_{out}}$ holding $b'_{_{_{\mathrm{BI}}}}\!\!(r_{_{0}}) = 1$. 

{\bf New magnetically charge T-WH in EGB Born-Infeld theory with variable GB coefficient.} 
Now, we want to focus on the limit case $\sigma_{_{0}}=0$ (for all $\beta$, $q$). For this limit case the metric \eqref{EsGB_BI}, with parameters satisfying \eqref{ell_imag}, preserves its T-WH interpretation. But now the 
scalar field \eqref{phiBI} becomes zero, whereas $\boldsymbol{f}(\phi)\Big|_{\sigma_{_{\!_{0}}}=0}$ becomes a nontrivial well-defined function of $r$. 
Therefore, for this case, the metric becomes an exact purely magnetic solution of an EGB Born-Infeld field theory 
with variable GB coefficient, 
$\alpha_{_{GB}}(r) =\int^{r}_{r_{_{0}}} \phi'(\tilde{r})\dot{\boldsymbol{f}}(\tilde{r})d\tilde{r}$, where $\phi'(r)\dot{\boldsymbol{f}}(r )$ is a well-defined function of the radial coordinate, given by 
\begin{equation}
\phi'(r)\dot{\boldsymbol{f}}(r)=\frac{2}{ \left( r^{2} \!-\! \sqrt{ r^{4} \!+\! \frac{q^{2}}{2\beta^{^{2}}} } ~\right)\!\!\sqrt{ 
4 \!+\! 32\beta^{^{2}}\!\!\left( r^{2}\!-\!\sqrt{ r^{4} \!+\! \frac{q^{2}}{2\beta^{^{2}}} } ~ \right) } 
}\!\!\int^{r}_{r_{_{0}}}\frac{ \chi \left( \chi^{2} \!-\! \sqrt{ \chi^{4} \!+\! \frac{q^{2}}{2\beta^{^{2}}} }  \right)}{  \sqrt{
\left(\chi^{4} \!+\! \frac{q^{2}}{2\beta^{^{2}}} \right) \!\! \left[ 4 \!+\! 32\beta^{^{2}}\!\!\left( \chi^{2}\!-\!\sqrt{ \chi^{4} + \frac{q^{2}}{2\beta^{^{2}}} } ~ \right) \right] ~ } 
}
d\chi. 
\end{equation}

\section{Conclusion}
In this work, we have presented a method to obtain magnetically charged, ultrastatic and spherically symmetric spacetime solutions of the EsGB-$\mathcal{L}(\mathcal{F})$ theory whose action is given by Eq. \eqref{actionL}, with a sGB coupling function $\boldsymbol{f}(\phi)$ constructed from an arbitrary electromagnetic Lagrangian density $\mathcal{L}(\mathcal{F})$. 
Using this method, considering vanishing electromagnetic Lagrangian $\mathcal{L}(\mathcal{F})=0$,
a theorem which discards the existence of ultrastatic,  
spherically symmetric and asymptotically flat traversable wormholes  
in pure EsGB theories [with $\boldsymbol{f}(\phi)\neq$ constant], has been proved.  
Using the Maxwell's electromagnetic theory $\mathcal{L}_{_{_{\mathrm{LED}}}} \!=\! \mathcal{F}$  
yields the  magnetically charged Ellis-Bronnikov EsGB Maxwell wormhole
\eqref{EsGB_maxwell} which can be interpreted as the magnetic dual of the purely electric solution derived in  Ref. \cite{Canate2019}.  
Additionally two novel magnetically charged ultrastatic T-WH solutions, 
with metrics \eqref{EsGB_EH} and \eqref{EsGB_BI}, respectively associated with the nonlinear electrodynamics models of Euler-Heisenberg (in the approximation of the weak-field limit) $\mathcal{L}_{_{_{\mathrm{EH}}}}\!=\!\mathcal{L}_{_{_{\mathrm{LED}}}} + \gamma \mathcal{F}^{2}\!/2$, and Born-Infeld $\mathcal{L}_{_{_{\mathrm{BI}}}} \!=\! -4\beta^{2} + 4\beta^{2} \sqrt{ 1 + \mathcal{F}\!/\!(2\beta^{2})~}$, have been constructed. 
Both solutions in the limit of weak electromagnetic field  
({\it i.e.} as $\mathcal{F}\approx0$) are reduced to 
the magnetically charged Ellis-Bronnikov EsGB Maxwell wormhole metric. 
Moreover, in each solution,  
the limiting cases, (i) absence of magnetic charge $q=0$ and (ii) vanishing of scalar charge $\sigma_{_{0}}=0$, were analyzed. In the former case the magnetic field disappears $\mathcal{F}=0$, 
the quantity $\boldsymbol{f}(\phi)$ becomes a constant function and the respective line elements reduce to the Ellis-Bronnikov wormhole metric as long as the scalar charge becomes imaginary $\sigma_{_{0}}\in\mathbb{I}$.  
In the latter case the T-WHs are kept open by the $\boldsymbol{f}$GB curvature with no need of exotic matter.
Thus, for the first time, the construction of T-WH geometries supported by viable models of nonlinear electrodynamics,  a real scalar field having positive kinetic term and $\boldsymbol{f}$GB curvature, has been discussed. 
In a forthcoming paper shall be determined the quasinormal modes, the corresponding Penrose diagrams, 
as well as we shall explore the stability of these novel T-WHs. In addition, in view of the results obtained in this work, it may also be interesting to investigate whether the stability of the wormhole can be guaranteed 
by a suitable choice of $\mathcal{L}(\mathcal{F})$.

\textbf{Acknowledgments}:
P. C. acknowledges financial support from Facultad de Ciencias Exactas y Naturales, Universidad Surcolombiana,
C\'odigo de Financiamiento 4132.

\appendix 
\section{\bf Null and weak energy conditions in EsGB-$\boldsymbol{\mathcal{L}(\mathcal{F})}$}
For an energy-momentum tensor $T_{\mu\nu}$, the null energy condition (NEC) stipulates that for every null vector, $n^{\alpha}$, yields $T_{\mu\nu}n^{\mu}n^{\nu}\geq0$.
Following \cite{WEC}, for a diagonal energy-momentum tensor $(T_{\alpha\beta})=\mathrm{diag} \left( T_{tt},T_{rr},T_{\theta\theta},T_{\varphi\varphi} \right)$, which can be
conveniently written as 
\begin{equation}\label{diagonalEab}
T_{\alpha}{}^{\beta} = - \rho_{t} \hskip.05cm \delta_{\alpha}{}^{t}\delta_{t}{}^{\beta} + P_{r} \hskip.05cm \delta_{\alpha}{}^{r}\delta_{r}{}^{\beta} + P_{\theta} \hskip.05cm \delta_{\alpha}{}^{\theta}\delta_{\theta}{}^{\beta} + P_{\varphi} \hskip.05cm \delta_{\alpha}{}^{\varphi}\delta_{\varphi}{}^{\beta} 
\end{equation}
 where $\rho_{t}$ may be interpreted as the rest energy density of the matter,  
 whereas $P_{r}$, $P_{\theta}$ and $P_{\varphi}$ are respectively the pressures along the $r$, $\theta$ and $\varphi$ directions. In terms of (\ref{diagonalEab}) the NEC implies
\begin{equation}\label{NEC_Eab}
\rho_{t} + P_{a} \geq0 \quad\quad\textup{with}\quad\quad  a = \{ r, \theta, \varphi \}.
\end{equation}
The weak energy condition (WEC) states that for any timelike vector $\boldsymbol{k} = k^{\mu}\partial_{\mu}$, (i.e., $k_{\mu}k^{\mu}<0$), the energy-momentum tensor  obeys the inequality
$T_{\mu\nu}k^{\mu}k^{\nu} \geq 0$, which means that the local energy density $\rho_{\!_{_{loc}}}= T_{\mu\nu}k^{\mu}k^{\nu}$ as measured by any observer with timelike vector 
$\boldsymbol{k}$ is a non-negative quantity.  
For an energy-momentum tensor of the form (\ref{diagonalEab}), the WEC will be satisfied if and only if
\begin{equation}\label{WEC}
\rho_{t} = - T_{t}{}^{t} \geq0, \quad\quad\quad\quad \rho_{t} + P_{a} \geq0 \quad\textup{with}\quad  a = \{ r, \theta, \varphi \}.
\end{equation}

%
\begin{center}
{\bf 1. NEC and WEC for a massless scalar field}
\end{center}

For static and spherically symmetric spacetime metric, identifying (\ref{SF_T}) with (\ref{diagonalEab}), and using (\ref{EttyErr}), yields
\begin{eqnarray}
&& 8\pi (\rho_{t})\!_{_{_{S \! F }}} = - 8\pi (P_{\theta})\!_{_{_{S \! F }}} = - 8\pi (P_{\varphi})\!_{_{_{S \! F }}} = \frac{1}{ 4 } e^{ -B} \phi'^{2}  \label{Energ}   
\\ 
&& 8\pi (P_{r})\!_{_{_{S \! F }}} = \frac{1}{ 4 } e^{ -B} \phi'^{2}
\end{eqnarray}
since $(\rho_{t})\!_{_{_{S \! F }}} + (P_{a})\!_{_{_{S \! F }}}=0$ for all $a=\theta$, $\varphi$, in this geometry the  energy-momentum tensor $(E_{\alpha}{}^{\beta})\!_{_{_{S \! F }}}$ satisfies the NEC,  Eq. \eqref{NEC_Eab}, if  
\begin{equation}\label{NEC_SF}
8\pi(\rho_{t})\!_{_{_{S \! F }}} + 8\pi (P_{r})\!_{_{_{S \! F }}}  = \frac{1}{ 2 } e^{ -B} \phi'^{2} \geq0.
\end{equation}
Since (\ref{Energ}), in this case the fulfillment of (\ref{NEC_SF}) implies $(\rho_{t})\!_{_{_{S \! F }}}>0$. 
Hence, in this case, NEC holds if and only if WEC holds.
%
\begin{center}
{\bf 2. NEC and WEC for the linear/nonlinear electrodynamics: Pure-magnetic case}
\end{center}

Using (\ref{NLED_EM}) and (\ref{E_nled}) yields 
%
\begin{eqnarray}
8\pi (\rho_{t})\!_{_{_{N \! L \! E \! D}}} = - 8\pi (P_{r})\!_{_{_{N \! L \! E \! D}}} =  2\mathcal{L},   
\quad\quad\quad 8\pi (P_{\theta})\!_{_{_{N \! L \! E \! D}}} = 8\pi (P_{\varphi})\!_{_{_{N \! L \! E \! D}}} =  2(2\mathcal{F}\mathcal{L}_{_{\mathcal{F}}} - \mathcal{L}). 
\end{eqnarray}
Since $\rho_{t} + P_{r}=0$, the tensor $(E_{\alpha}{}^{\beta})\!_{_{_{N \! L \! E \! D}}}$ satisfies the NEC if
\begin{equation}\label{NEC_NLED}
8\pi(\rho_{t})\!_{_{_{N \! L \! E \! D}}} + 8\pi (\rho_{\theta})\!_{_{_{N \! L \! E \! D}}}   = 8\pi(\rho_{t})\!_{_{_{N \! L \! E \! D}}} + 8\pi(\rho_{\varphi})\!_{_{_{N \! L \! E \! D}}}  =  4\mathcal{F}\mathcal{L}_{_{\mathcal{F}}}\geq0.
\end{equation}
In addition to (\ref{NEC_NLED}) if    
\begin{equation}\label{WEC_NLED}
8\pi (\rho_{t})\!_{_{_{N \! L \! E \! D}}} =  2\mathcal{L}\geq0
\end{equation}
the WEC is satisfied.  
%
%

\section{\bf Magnetically charge ultrastatic and spherically symmetric solution of the Einstein-Gauss-Bonnet theory (with variable Gauss-Bonnet coefficient) coupled to $\boldsymbol{\mathcal{L}(\mathcal{F})}$ electrodynamics }

According to Eq. \eqref{phi_sol} the trivial case of vanishing scalar field, $\phi(r)=0$, is obtained when setting $\sigma_{_{0}}=0$, whereas for this  case the line element \eqref{UltraS_Solution}, with shape function \eqref{Fsol}, takes the nontrivial form
\begin{equation}\label{FsolTr}
ds^{2} =  - dt^{2} + \frac{dr^{2}}{ 1 - 2r^{2}\mathcal{L}(\mathcal{F}) }  + r^{2}(d\theta^{2}  + \sin^{2}\theta d\varphi^{2}), \quad\quad \textup{with} \quad\quad \mathcal{F} =  \frac{q^{2}}{r^{4}}.
\end{equation}
However,  $\sigma_{_{0}}\rightarrow0$ also implies that $\phi' = 0$, $\phi'' = 0$, 
$\dot{\boldsymbol{f}}(\phi) \rightarrow \infty$, and $\ddot{\boldsymbol{f}}(\phi) \rightarrow \infty$. Despite this, the quantities $\dot{\boldsymbol{f}}(\phi)$ and $\ddot{\boldsymbol{f}}(\phi)$ appear in the EsGB-$\mathcal{L}(\mathcal{F})$ field equations as multiplicative factors to derivatives of the scalar field, i.e.,  $\phi'\dot{\boldsymbol{f}}(\phi)$, $\phi''\dot{\boldsymbol{f}}(\phi)$ and $\phi'^{2}\ddot{\boldsymbol{f}}(\phi)$, and these (when $\sigma_{_{0}} = 0$) are nonzero and regular functions in the spacetime domain, and are given by 
\begin{eqnarray}
 \phi'(r)\dot{\boldsymbol{f}}(r)  
 &=& \frac{r}{4b(r)}\left( 1 - \frac{b(r)}{r}\right)^{-\frac{1}{2}} 
 \mathcal{P}(r) 
 \label{C3phif}\\
 \phi''(r)\dot{\boldsymbol{f}}(r)  
 &=& \frac{[3b(r) - 4r + rb'(r)]r^{^{\frac{1}{2}}}}{8[r-b(r)]^{^{\frac{3}{2}}}b(r) } \mathcal{P}(r)  
 \label{C4phif}\\
 \phi'^{2}(r)\ddot{\boldsymbol{f}}(r)  
 &=& \frac{r}{4b(r)} \left( 1 - \frac{b(r)}{r}\right)^{-\frac{1}{2} }\bigg\{ \mathcal{P}'(r) - \left[\ln\left(\frac{b(r)}{r^{3}}\right)\right]' \mathcal{P}(r) 
\bigg\} \label{C5phif}
\end{eqnarray}
where we use the auxiliary function
\begin{equation}\label{C2phif}
    \mathcal{P}(r) = \int^{r}_{r_{_{0}}}  b(\chi)\left[\ln\left(\frac{b(\chi)}{\chi}\right)\right]' \left( 1 - \frac{b(\chi)}{\chi}\right)^{-\frac{1}{2}}d\chi.
\end{equation}
The quantities (\ref{C3phif}), (\ref{C4phif}) and (\ref{C5phif}) are not independent; they satisfy
\begin{equation}
(\phi'\dot{\boldsymbol{f}})' = \phi''\dot{\boldsymbol{f}} + \phi'^{2}\ddot{\boldsymbol{f}}.    
\end{equation}
Moreover, for this case the scalar field $\phi$ and the coupling function $\boldsymbol{f}$ are decoupled, in such a way that $\boldsymbol{f}$ as a function of radial coordinate takes the form
\begin{equation}\label{C2fdr}
\boldsymbol{f}(r) = \int^{r}_{r_{_{0}}} \frac{\tilde{r}}{4b(\tilde{r})}\left( 1 - \frac{b(\tilde{r})}{\tilde{r}}\right)^{-\frac{1}{2}} \mathcal{P}(\tilde{r})d\tilde{r}.
\end{equation}
Summarizing, we can say that when $\sigma_{_{0}}=0$, the scalar field becomes trivial, $\phi(r)=0$. However, the line element \eqref{FsolTr}, together with the linear/nonlinear electromagnetic Lagrangian density $\mathcal{L}(\mathcal{F})$, scalar field and coupling function given respectively by \eqref{phi_sol} and \eqref{df_sol}, is still a  solution of EsGB-$\mathcal{L}(\mathcal{F})$.  In this case the parameter $q$ in the line element \eqref{FsolTr} represents the magnetic charge and the ultrastatic solution is only supported by the source-free magnetic field and the $\boldsymbol{f}$GB curvature. \\ 
Alternatively, the metric \eqref{FsolTr} can be also interpreted as an exact purely magnetic solution of an of EGB-$\mathcal{L}(\mathcal{F})$ theory with variable GB coefficient $\alpha_{_{GB}}(r)=\boldsymbol{f}(r)$ given by Eq. \eqref{C2fdr}; hence, $\alpha'_{_{GB}}(r) =  \phi'(r)\dot{\boldsymbol{f}}(r)$ given by \eqref{C3phif}, whereas $\alpha''_{_{GB}}(r) = \phi''(r)\dot{\boldsymbol{f}}(r) + \phi'^{2}(r)\ddot{\boldsymbol{f}}(r)$ given by the sum of Eqs. \eqref{C4phif} and \eqref{C5phif}.

\end{document}